\documentclass[aps,pre,groupedaddress,showpacs,amsmath,eqsecnum]{revtex4}
\usepackage{amsfonts}
\usepackage{amssymb}
\usepackage{amsmath}
\usepackage{graphicx}
\usepackage{amsbsy}
\usepackage{color}

\newcommand{\bq}{\begin{eqnarray}}
\newcommand{\eq}{\end{eqnarray}}
\newcommand{\bqn}{\begin{eqnarray*}}
\newcommand{\eqn}{\end{eqnarray*}}

\begin{document}

\preprint{}
\title{Fluids of spherical molecules \\
with dipolar-like nonuniform adhesion. \\
An analytically solvable anisotropic model.}
\author{Domenico Gazzillo, Riccardo Fantoni and Achille Giacometti}
\affiliation{Dipartimento di Chimica Fisica, Universit\`a di Venezia, S. Marta DD 2137,
I-30123 Venezia, Italy}

\begin{abstract}
We consider an anisotropic version of Baxter's model of `sticky hard
spheres', where a nonuniform adhesion is implemented by adding, to an
isotropic surface attraction, an appropriate `dipolar sticky' correction
(positive or negative, depending on the mutual orientation of the
molecules). The resulting nonuniform adhesion varies continuously, in such a
way that in each molecule one hemisphere is `stickier' than the other.

We derive a complete analytic solution by extending a formalism [M.S.
Wertheim, J. Chem. Phys. \textbf{55}, 4281 (1971) ] devised for dipolar hard
spheres. Unlike Wertheim's solution which refers to the `mean spherical
approximation', we employ a \textit{Percus-Yevick closure with orientational
linearization}, which is expected to be more reliable.

We obtain analytic expressions for the orientation-dependent pair
correlation function $g\left( 1,2\right) $. Only one equation for a
parameter $K$ has to be solved numerically. We also provide very accurate
expressions which reproduce $K$ as well as some parameters, $\Lambda _{1}$
and $\Lambda _{2}$, of the required Baxter factor correlation functions with
a relative error smaller than $1\%$. We give a physical interpretation of
the effects of the anisotropic adhesion on the $g\left( 1,2\right) $.

The model could be useful for understanding structural ordering in complex
fluids within a unified picture.
\end{abstract}

\date{\today}
\startpage{1}
\maketitle

%

\section{INTRODUCTION}

Anisotropy of molecular interactions plays an important role in many
physical, chemical and biological processes. Attractive forces are
responsible for the tendency toward particle association, while the
directionality of the resulting bonds determines the geometry of the
resulting clusters. Aggregation may thus lead to very different structures:
in particular, chains, globular forms, and bi- or three-dimensional
networks. Understanding the microscopic mechanisms underlying such phenomena
is clearly very important both from a theoretical and a technological point
of view. Polymerization of inorganic molecules, phase behaviour of
non-spherical colloidal particles, building up of micelles, gelation,
formation of $\alpha $-helices from biomolecules, DNA-strands, and other
ordered structures in living organisms, protein folding and crystallization,
self-assembly of nanoparticles into composite objects designed for new
materials, are all subjects of considerable interest, belonging to the same
class of systems with anisotropic interactions.

Modern studies on these complex systems strongly rely upon computer
simulations, which have provided a number of useful information about many
properties of molecular fluids.

Nevertheless, analytic models with explicit expressions for structural and
thermodynamic properties still represent an irreplaceable tool, in view of
their ability of capturing the essential features of the investigated
physical systems.

At the lowest level in this hierarchy of minimal models on assembling
particles lies the problem of the formation of linear aggregates, from
dimers \cite{Spinozzi02,Giacometti05} up to polymer chains. This topic has
been extensively investigated, through both computer simulations and
analytical methods. In the latter case a remarkable example is Wertheim's
analytic solution of the \textit{mean spherical approximation} (MSA)
integral equation for dipolar hard spheres (DHS), i.e. hard spheres (HS)
with a point dipole at their centre \cite{Wertheim71} (hereafter referred to
as I). For the DHS model, several studies predict chain formation, whereas
little can be said about the existence of a fluid-fluid coexistence line,
since computer simulations and mean field theories provide contradictory
results \cite{Weis93,Leeuwen93,Sear96,Camp00,Tlusty00}. On the other hand,
for mesoscopic fluids the importance of combining \textit{short-ranged}
anisotropic attractions and repulsions has been well established \cite%
{Gazzillo06,Fantoni07}, and hence the long-range of the dipolar interaction
is less suited for the mesoscopic systems considered here, at variance with
their atomistic counterpart.

The aim of the present paper is to address both the above points, by
studying a model with anisotropic surface adhesion that is amenable to an
analytical solution, within an approximation which is expected to be valid
at significant experimental regimes.

In the isotropic case, the first model with `surface adhesion' was
introduced long time ago by Baxter \cite{Baxter68,Baxter71}. The interaction
potential of these `sticky hard spheres' (SHS) includes a HS repulsion plus
a spherically symmetric attraction, described by a square-well (SW) which
becomes infinitely deep and narrow, according to a limiting procedure
(Baxter's sticky limit) that keeps the second virial coefficient finite.

Possible anisotropic variations include `sticky points'\ \cite%
{Sciortino05,Bianchi06,Michele06,Lomakin99,Starr03,Zhang04,Glotzer04a,Glotzer04b,Sciortino07}%
, `sticky patches'\ \cite%
{Jackson88,Ghonasgi95,Sear99,Mileva00,Kern03,Zhang05,Fantoni07} and, more
recently, `Gaussian patches'\ \cite{Wilber06,Doye07}. The most common
version of patchy sticky models refers to HS with one or more `uniform
circular patches', all of the same species. This kind of patch has a
well-defined circular boundary on the particle surface, and is always
attractive, with an `uniform' strength of adhesion, which does not depend on
the contact point within the patch \cite{Jackson88}.

In the present paper we consider a `dipolar-like' SHS model, where the sum
of a uniform surface adhesion (isotropic background) plus an appropriate
dipolar sticky correction -- which can be both positive or negative,
depending on the orientations of the particles -- yields a nonuniform
adhesion. Although the adhesion varies continuously and no discontinuous
boundary exists, the surface of each molecule may be regarded as formed by
two hemispherical `patches' (colored red and blue, respectively, in the
online Figure 1). One of these hemispheres is `stickier' than the other, and
the entire molecular surface is adhesive, but its stickiness is nonuniform
and varies in a dipolar fashion. By varying the dipolar contribution, the
degree of anisotropy can be changed, in such a way that the total sticky
potential can be continuously tuned from very strong attractive strength
(twice the isotropic one) to vanishing adhesion (HS limit). The physical
origin of this model may be manifold (non-uniform distribution of surface
charges, or hydrophobic attraction, or other physical mechanisms), one
simple realization being as due to an `extremely screened' attraction. The
presence of a solvent together with a dense ionic atmosphere could induce
any electrostatic interaction to vanish close to the molecular surface, and
-- in the idealized sticky limit -- to become \textit{truncated }exactly at
contact.

For this model, we solve analytically the molecular Ornstein-Zernike (OZ)
integral equation, by using a truncated \textit{Percus-Yevick} (PY)
approximation, \textit{with orientational linearization }(PY-OL), since it
retains only the lowest order terms in the expansions of the correlation
functions in angular basis functions. This already provides a clear
indication of the effects of anisotropy on the adhesive adhesion.

The idea of an anisotropic surface adhesion is not new. In a series of
papers on hydrogen-bonded fluids such a water, Blum and co-workers \cite%
{Cummings86,Wei88,Blum90} already studied models of spherical molecules with
anisotropic pair potentials, including both electrostatic multipolar
interactions and sticky adhesive terms of multipolar symmetry. Within
appropriate closures, these authors outlined the general features of the
analytic solutions of the OZ equation by employing a very powerful formalism
based upon expansions in rotational invariants. In particular, Blum,
Cummings and Bratko \cite{Blum90} obtained an analytic solution within a
mixed MSA/PY closure (extended to mixtures by Protsykevich \cite%
{Protsykevich03}) for molecules which have surface adhesion of dipolar
symmetry and at most dipole-dipole interactions. From the physical point of
view, our model -- with `dipolar-like' adhesion resulting from the sum of an
isotropic plus a dipolar term -- is different and more specifically
characterized with respect to the one of Ref. [32], whose adhesion has a
simpler, strictly `dipolar', symmetry. From the mathematical point of view,
however, the same formalism employed by Blum \textit{et al.} \cite{Blum90}
could also be applied to our model. Unfortunately, the solution given in
Ref. [32] is not immediately usable for the actual computation of
correlation functions, since the explicit determination of the parameters
involved in their analytical expressions is lacking.

In the present paper we adopt a simpler solution method, by extending the
elegant approach devised by Wertheim for DHS within the MSA closure \cite%
{Wertheim71}, and, most importantly, we aim at providing a \textit{complete}
analytic solution -- including the determination of all required parameters
-- within our PY-OL approximation.

The paper is organized as follows. Section II defines the model. In Section
III we recall the molecular OZ integral equation and the basic formalism. In
Section IV we present the analytic solution. Numerical exact results for
some necessary parameters, as well as very accurate analytic approximations
for them, will be shown in Section V. Some preliminary plots illustrating
the effects of the anysotropic adhesion on the local structure are reported
in Section VI. Phase stability is breafly discussed in Section VII, while
final remarks and conclusions are offered in Section VIII.\bigskip

\bigskip

\section{HARD SPHERES WITH\ ADHESION OF\ DIPOLAR-LIKE\ SYMMETRY}

Let the symbol $i\equiv \left( \mathbf{r}_{i},\Omega _{i}\right) $ (with $%
i=1,2,3,\ldots $) denote both the position $\mathbf{r}_{i}$ of the molecular
centre and the orientation $\Omega _{i}$ of molecule $i$; for linear
molecules, $\Omega _{i}\equiv \left( \theta _{i},\varphi _{i}\right) $
includes the usual polar and azimuthal angles. Translational invariance for
uniform fluids allows to write the dependence of the pair correlation
function $g\left(1,2\right)$ as 
\begin{equation*}
(1,2)=(\mathbf{r}_{12},\Omega _{1},\Omega _{2})=(r,\Omega _{1},\Omega _{2},%
\widehat{\mathbf{r}}_{12})=(r,\Omega _{1},\Omega _{2},\Omega _{r}),
\end{equation*}%
with $\mathbf{r}_{12}=\mathbf{r}_{2}-\mathbf{r}_{1}$, $r=|\mathbf{r}_{12}|$,
and $\Omega _{r}$ being the solid angle associated with $\widehat{\mathbf{r}}%
_{12}=\mathbf{r}_{12}/r.$

In the spirit of Baxter's isotropic counterpart \cite{Baxter68,Gazzillo04},
our model is defined by the Mayer function given by 
\begin{equation}
f^{\mathrm{SHS}}(1,2)=f^{\mathrm{HS}}(r)+t\ \epsilon (1,2)\ \sigma \delta
\left( r-\sigma \right) ,  \label{eq2}
\end{equation}%
where $f^{\mathrm{HS}}(r)=\Theta \left( r-\sigma \right) -1$ is its HS
counterpart, $\Theta $ is the Heaviside step function ($\Theta (x<0)=0$, $%
\Theta (x>0)=1$) and $\delta \left( r-\sigma \right) $ the Dirac delta
function, which ensures that the adhesive interaction occurs only at contact
($\sigma $ being the hard sphere diameter). An appropriate limit of the
following particular square well potential of width $R-\sigma $ 
\begin{equation*}
\Phi ^{\mathrm{SW}}\left( 1,2\right) =\left\{ 
\begin{array}{ccc}
+\infty \text{ \ \ \ \ \ \ \ \ \ \ \ \ \ \ \ \ \ \ \ \ \ \ \ \ \ \ \ \ \ \ \
\ \ \ } &  & 0<r<\sigma \\ 
-k_{B}T\ \ln \left[ 1+t\ \epsilon (1,2)\ \frac{\sigma }{R-\sigma }\ \ \right]
&  & \sigma <r<R \\ 
0\text{ \ \ \ \ \ \ \ \ \ \ \ \ \ \ \ \ \ \ \ \ \ \ \ \ \ \ \ \ \ \ \ \ \ \
\ \ \ } &  & \text{ \ \ \ \ \ \ \ }r>R\text{ ,}%
\end{array}%
\right.
\end{equation*}%
can be shown to lead to Eq. (\ref{eq2}).

The angular dependence is buried in the angular factor%
\begin{equation}
\epsilon (1,2)=1+\alpha D(1,2),
\end{equation}%
including the dipolar function%
\begin{equation*}
D(1,2)=D(\Omega _{1},\Omega _{2},\Omega _{r})=3(\mathbf{u}_{1}\cdot \hat{%
\mathbf{r}})(\mathbf{u}_{2}\cdot \hat{\mathbf{r}})-\mathbf{u}_{1}\cdot 
\mathbf{u}_{2}
\end{equation*}%
which stems from the dipole-dipole potential $\phi ^{\mathrm{dip-dip}%
}(1,2)=-\mu ^{2}D(1,2)/r^{3}$ ($\mu $ is the magnitude of the dipole moment)
and is multiplied by the tunable \textit{anisotropy parameter} $\alpha $. In
the isotropic case, $\alpha =0$, one has $\epsilon (1,2)=1$. Here and in the
following, $\hat{\mathbf{r}}$ coincides with $\hat{\mathbf{r}}_{12}=-\hat{%
\mathbf{r}}_{21}$ , while $\mathbf{u}_{i\text{ }}$is the versor attached to
molecule $i$ (drawn as yellow arrow in Figure 1) which completely determines
its orientation $\Omega _{i}$. Note the symmetry $D(2,1)=D(1,2)$.

The condition $\epsilon (1,2)\geq 0$ must be enforced in order to preserve a
correct definition of the sticky limit, ensuring that the total sticky
interaction remains attractive for all orientations, and the range of
variability $-2\leq D(1,2)\leq 2$ yields the limitation $0\leq \alpha \leq 
\frac{1}{2}$ on the anisotropy degree. The stickiness parameter $t$ -- equal
to $\left( 12\tau \right) ^{-1}$ in Baxter's original notation \cite%
{Baxter68} -- measures the strength of surface adhesion relatively to the
thermal energy $k_{B}T$ ($k_{B}$ being the Boltzmann constant, $T$ the
absolute temperature) and increases with decreasing temperature.

If we adopt an `inter-molecular reference frame' (with both polar axis and
cartesian $z$-axis taken along $\mathbf{r}_{12}$), then the cartesian
components of $\hat{\mathbf{r}}$ and $\mathbf{u}_{i}$ are $(0,0,1)$ and $%
(\sin \theta _{i}\cos \varphi _{i}$, $\sin \theta _{i}\sin \varphi _{i}$, $%
\cos \theta _{i})$, respectively, and thus%
\begin{equation}
D(1,2)=2\cos \theta _{1}\cos \theta _{2}-\sin \theta _{1}\sin \theta
_{2}\cos \left( \varphi _{1}-\varphi _{2}\right) .  \label{eq4b}
\end{equation}

The strength of adhesion between two particles $1$ and $2$ at contact
depends -- in a continuous way -- on the relative orientation of $\mathbf{u}%
_{1}$ and $\mathbf{u}_{2}$ as well as on the versor $\widehat{\mathbf{r}}%
_{12}$ of the intermolecular distance. We shall call \textit{parallel} any
configuration with $\mathbf{u}_{1}\cdot \mathbf{u}_{2}=1$, while \textit{%
antiparallel} configurations are those with $\mathbf{u}_{1}\cdot \mathbf{u}%
_{2}=-1$ (see Figure 1). For all configurations with $D(1,2)>0$, the
anisotropic part of adhesion is attractive and adds to the isotropic one.
Thus, the surface adhesion is maximum, and larger than in the isotropic
case, when $\mathbf{u}_{1}=\mathbf{u}_{2}=$ $\widehat{\mathbf{r}}_{12}$ and
thus $\epsilon (1,2)=1+2\alpha $ (head-to-tail parallel configuration, shown
in Figure 1b). On the contrary, when $D(1,2)<0$ the anisotropic contribution
is repulsive and subtracts from the isotropic one, so that the total sticky
interaction still remains attractive. Then, the stickiness is minimum, and
may even vanish for $\alpha =1/2$, when $\mathbf{u}_{1\text{ }}=-$ $\mathbf{u%
}_{2}=\widehat{\mathbf{r}}_{12}$ and thus $\epsilon (1,2)=1-2\alpha $
(head-to-head or tail-to-tail antiparallel configurations, reported in
Figure 1c). The intermediate case of \textit{orthogonal }configuration ($%
\mathbf{u}_{2}$ perpendicolar to $\mathbf{u}_{1}$) corresponds to $D(1,2)=0,$
which is equivalent to the isotropic SHS interaction.

It proves convenient to `split' $f^{\mathrm{SHS}}(1,2)$ as 
\begin{equation}
\text{\ }f^{\mathrm{SHS}}(1,2)=f_{0}(r)+f_{\mathrm{ex}}(1,2),  \label{eq6a}
\end{equation}%
\begin{equation}
\left\{ 
\begin{array}{c}
f_{0}(r)=f^{\mathrm{HS}}(r)+t\ \sigma \delta \left( r-\sigma \right) \equiv
f^{\mathrm{isoSHS}}(r) \\ 
f_{\mathrm{ex}}(1,2)=\left( \alpha t\right) \ \sigma \delta \left( r-\sigma
\right) \ D(1,2),\text{ \ \ \ \ \ \ \ \ }%
\end{array}%
\right.  \label{eq6b}
\end{equation}%
where the spherically symmetric $f_{0}(r)$ corresponds to the `reference'
system with isotropic background adhesion, while $f_{\mathrm{ex}}(1,2)$ is
the orientation-dependent `excess' term.

We remark that, as shown in Ref. I (see also Table I in Appendix A of the
present paper), convolutions of $f^{\mathrm{SHS}}$-functions generate
correlation functions with a more complex angular dependence. Therefore, in
addition to $D(1,2)$, it is necessary to consider also 
\begin{equation}
\Delta (1,2)=\mathbf{u}_{1}\cdot \mathbf{u}_{2}\text{ }=\cos \theta _{1}\cos
\theta _{2}+\sin \theta _{1}\sin \theta _{2}\cos \left( \varphi _{1}-\varphi
_{2}\right) ,\ 
\end{equation}%
where the last equality holds true in the inter-molecular frame. The limits
of variation for $\Delta (1,2)$ are clearly $-1\leq \Delta (1,2)\leq 1$.

\section{BASIC FORMALISM}

This section, complemented by Appendix A, presents the main steps of \
Wertheim's formalism, as well as its extension to our model.

\subsection{Molecular Ornstein-Zernike equation}

The \textit{molecular OZ integral equation} for a pure and homogeneous fluid
of molecules interacting via non-spherical pair potentials is

\begin{equation}
h(1,2)=c(1,2)+\rho \int d\mathbf{r}_{3}\ \left\langle \ c(1,3)\ h(3,2)\
\right\rangle _{\Omega _{3}}\ ,  \label{oz4}
\end{equation}%
where $h(1,2)$ and $c(1,2)$ are the total and direct correlation functions,
respectively, $\rho $ is the number density, and $g(1,2)=1+h(1,2)$ is the
pair distribution function \cite{Friedman85,Lee88,Hansen06}. Moreover, the
angular brackets with subscript $\Omega $ denote an average over the
orientations, i.e. $\left\langle \cdots \right\rangle _{\Omega }=\left( 4\pi
\right) ^{-1}\int d\Omega \ \cdots .$

The presence of convolution makes convenient to Fourier transform (FT) this
equation, by integrating with respect to the space variable $\mathbf{r}$
alone according to

\begin{equation}
\widehat{F}\left( \mathbf{k},\Omega _{1},\Omega _{2}\right) =\int d\mathbf{r}%
\ F(\mathbf{r},\Omega _{1},\Omega _{2})\ \exp (i\mathbf{k\cdot r}).
\label{oz5}
\end{equation}%
The $\mathbf{r}$-space convolution becomes a product in $\mathbf{k}$-space,
thus leading to%
\begin{equation}
\widehat{h}(\mathbf{k},\Omega _{1},\Omega _{2})=\widehat{c}(\mathbf{k}%
,\Omega _{1},\Omega _{2})+\rho \ \left\langle \widehat{c}(\mathbf{k},\Omega
_{1},\Omega _{3})\ \widehat{h}(\mathbf{k},\Omega _{3},\Omega
_{2})\right\rangle _{\Omega _{3}}\ .  \label{oz6}
\end{equation}

As usual the OZ equation involves two unknown functions, $h$ and $c$, and
can be solved only after adding a closure, that is a second (approximate)
relationship among $c$, $h$ and the potential.

\subsection{Splitting of the OZ equation: reference and excess part}

The particular form of our potential, as defined by the Mayer function of
Eq. (\ref{eq2}), gives rise to a remarkable exact splitting of the original
OZ equation. Using diagrammatic methods \cite{Friedman85,Lee88,Hansen06} it
is easy to see that both $c$ and $h$ can be expressed as graphical series
containing the Mayer function $f$ as \textit{bond function}. If $\ f^{%
\mathrm{SHS}}=f_{0}+f_{\mathrm{ex}}$ is substituted into all graphs of the
above series, each diagram with $n$ $f$-bonds will generate $2^{n}$ new
graphs. In the cluster expansion of $c$, the sum of all graphs having only $%
f_{0}$-bonds will yield $c_{0}(r)=c^{\mathrm{isoSHS}}(r)$, i.e. the the
direct correlation function (DCF) of the reference fluid with isotropic
adhesion. On the other hand, all remaining diagrams have \textit{at least one%
} $f_{\mathrm{ex}}$-bond, whose expression is given by Eq. (\ref{eq6b}).
Thus, in the sum of this second subset of graphs it is possible to factorize 
$\alpha t$, and we can write 
\begin{equation}
\text{\ }c^{\mathrm{SHS}}(1,2)=c_{0}(r)+c_{\mathrm{ex}}(1,2),  \label{oz7a}
\end{equation}%
\begin{equation}
\text{ \ \ \ \ \ \ \ \ \ }\left\{ 
\begin{array}{c}
c_{0}(r)=c^{\mathrm{isoSHS}}(r),\text{\ \ \ \ \ \ \ \ \ \ \ \ \ \ \ \ \ \ }
\\ 
c_{\mathrm{ex}}(1,2)=\left( \alpha t\right) \ c^{\dagger }(1,2).\text{ \ \ \
\ \ \ \ \ \ \ }%
\end{array}%
\right.  \label{oz7b}
\end{equation}%
Similarly, for $h$ we get

\begin{equation}
\text{\ }h^{\mathrm{SHS}}(1,2)=h_{0}(r)+h_{\mathrm{ex}}(1,2),  \label{oz8a}
\end{equation}%
\begin{equation}
\text{ \ \ \ \ \ \ \ \ \ }\left\{ 
\begin{array}{c}
h_{0}(r)=h^{\mathrm{isoSHS}}(r),\text{\ \ \ \ \ \ \ \ \ \ \ \ \ \ \ \ \ \ }
\\ 
h_{\mathrm{ex}}(1,2)=\left( \alpha t\right) \ h^{\dagger }(1,2).\text{ \ \ \
\ \ \ \ \ \ \ }%
\end{array}%
\right.  \label{oz8b}
\end{equation}%
\ \ \ \ \ 

Note that this useful separation into reference and excess part may also be
extended to other correlation functions, such as $\gamma (1,2)\equiv
h(1,2)-c(1,2)$, $g(1,2)=1+h(1,2)$, and the `cavity' function $%
y(1,2)=g(1,2)/e(1,2)$. The function $\gamma $ coincides with the OZ
convolution integral, without singular $\delta $-terms. Similarly $y$ is
also `regular', and its exact expression reads $y\left( 1,2\right) =\exp %
\left[ \ \gamma \left( 1,2\right) +B(1,2)\ \right] $, where the `bridge'
function $B$ is defined by a complicated cluster expansion \cite%
{Friedman85,Lee88,Hansen06}.

From Eqs. (\ref{oz7a})-(\ref{oz8b}), which are merely a consequence of the
particular form of $f_{\mathrm{ex}}$ in the splitting of $f^{\mathrm{SHS}}$,
one immediately sees that, if the anisotropy degree\textit{\ }$\alpha $
tends to zero, then%
\begin{equation}
\lim_{\alpha \rightarrow 0}c_{\mathrm{ex}}(1,2)=\lim_{\alpha \rightarrow
0}h_{\mathrm{ex}}(1,2)=\lim_{\alpha \rightarrow 0}y_{\mathrm{ex}}(1,2)=0.
\label{oz9}
\end{equation}

Note that the spherically symmetric parts $c_{0}$ and $h_{0}$ must be
related through the OZ equation for the reference fluid with isotropic
adhesion (\textit{reference OZ equation})

\begin{equation}
h_{0}(r)=c_{0}(r)+\rho \int d\mathbf{r}_{3}\ c_{0}(r_{13})\ h_{0}(r_{32}).\ 
\label{ozeq1}
\end{equation}%
Thus, substituting $c$ and $h$ of Eq. (\ref{oz4}) with $c_{0}+c_{\mathrm{ex}%
} $ and $h_{0}+h_{\mathrm{ex}}$, respectively, and subtracting Eq. (\ref%
{ozeq1}), we find that $c_{\mathrm{ex}}$ and $h_{\mathrm{ex}}$ must obey the
following relation

\begin{eqnarray*}
h_{\mathrm{ex}}(1,2) &=&c_{\mathrm{ex}}(1,2)+\rho \int d\mathbf{r}_{3}\ %
\left[ \ c_{0}(r_{13})\ \left\langle \ h_{\mathrm{ex}}(3,2)\ \right\rangle
_{\Omega _{3}}\right. \\
&&\left. +\left\langle \ c_{\mathrm{ex}}(1,3)\ \right\rangle _{\Omega
_{3}}h_{0}(r_{32})+\left\langle \ c_{\mathrm{ex}}(1,3)\ h_{\mathrm{ex}%
}(3,2)\ \right\rangle _{\Omega _{3}}\ \right] \ .
\end{eqnarray*}%
and when 
\begin{equation}
\left\langle \ c_{\mathrm{ex}}(1,3)\ \right\rangle _{\Omega
_{3}}=\left\langle \ h_{\mathrm{ex}}(3,2)\ \right\rangle _{\Omega _{3}}=0
\label{eq_cond}
\end{equation}%
the orientation-dependent excess parts $c_{\mathrm{ex}}$ and $h_{\mathrm{ex}%
} $ satisfy the equality 
\begin{equation}
h_{\mathrm{ex}}(1,2)=c_{\mathrm{ex}}(1,2)+\rho \int d\mathbf{r}_{3}\
\left\langle \ c_{\mathrm{ex}}(1,3)\ h_{\mathrm{ex}}(3,2)\ \right\rangle
_{\Omega _{3}}\ ,  \label{ozeq2}
\end{equation}%
which is decoupled from that of the reference fluid and may be regarded as
an OZ equation for the excess part (\textit{excess OZ equation}). As we
shall see, condition (\ref{eq_cond}) is satisfied in our scheme.

We stress that, in principle, the closures for Eq. (\ref{ozeq1}) and Eq. (%
\ref{ozeq2}), respectively, might be \textit{different}. In addition,
although the two OZ equations\ are decoupled, a suitably selected closure
might establish a relationship between $F_{0}$ and $F$ $(F=c,h).$

\subsection{Percus-Yevick closure with orientational linearization}

For hard-core fluids, $h$ and $c$ inside the core are given by 
\begin{equation}
\left\{ 
\begin{array}{ccc}
h(1,2)=-1\text{ \ \ \ \ \ \ \ \ \ \ \ \ \ } &  & \text{for \ }0<r<\sigma ,
\\ 
c(1,2)=-\left[ 1+\gamma (1,2)\right] &  & \text{ for \ }0<r<\sigma .%
\end{array}%
\right.  \label{oz6a}
\end{equation}%
At the same time, we have the following \textit{exact} relations 
\begin{eqnarray*}
h(1,2) &=&g(1,2)-1=e(1,2)y\left( 1,2\right) -1, \\
c(1,2) &=&f(1,2)\left[ 1+\gamma (1,2)\right] +e(1,2)\left[ y\left(
1,2\right) -1-\gamma (1,2)\right] .
\end{eqnarray*}

Since $c$, $h$ and $g$ are discontinuous for hard-core fluids and involve $%
\delta $-terms for sticky particles, it is more convenient to define
closures in terms of $y$ and $\gamma $, which are still continuous and
without $\delta $-singularities. The \textit{Percus-Yevick} approximation
for molecular fluids with orientation-dependent interactions corresponds to
assuming%
\begin{equation}
y^{\mathrm{PY}}\left( 1,2\right) =1+\gamma (1,2)\text{ \ \ \ everywhere,}
\label{w3a}
\end{equation}%
and thus, for the DCF,%
\begin{equation}
c^{\mathrm{PY}}\left( 1,2\right) =f(1,2)\ \left[ \ 1+\gamma (1,2)\ \right] 
\text{,}  \label{w3b}
\end{equation}%
which implies that $c$ vanishes beyond the range of the potential.

However, the dependence of $\gamma (1,2)$ on angles may still be very
complex. A possible procedure is to perform a series expansion of all
correlation functions in terms of an \textit{infinite} set of \textit{%
rotational invariants}, which are angular basis functions -- related to the
spherical harmonics -- having the property of rotational invariance valid
for homogeneous fluids \cite{Blum71}. Unfortunately, the full PY
approximation requires an infinite number of expansion coefficients for both 
$c(1,2)$ and $h(1,2)$. This approach is usually impracticable, but sometimes
even unnecessary, as it is possible that the most significant angular basis
functions are included in a small closed subset of that infinite set. Indeed
this happens, for instance, in the DHS model within the MSA \cite{Wertheim71}%
, where the set $\{1,\Delta ,D\}$ is the required subset. Although this does 
\textit{not} happen in our model, we shall argue that the same truncation is
sufficient due to the dipolar symmetry of the anisotropic adhesion.

Indeed, a natural assumption is that the only nonzero harmonics in $c(1,2)$
and $h(1,2)$ are those contained in $f(1,2)$ and those which can be obtained
from that set by convolution \cite{Cummings86}. Now, the angular basis
functions included in our $f$-bond are only $1$ and $D$, but the convolution
of two $f$-bonds involves the angular average of two $D^{\prime }$s, which
yields \cite{Wertheim71}%
\begin{equation*}
\ \left\langle \ D(\mathbf{k},\Omega _{1},\Omega _{3})\ D(\mathbf{k},\Omega
_{3},\Omega _{2})\ \right\rangle _{\Omega _{3}}=\frac{1}{3}\left[ D(\mathbf{k%
},\Omega _{1},\Omega _{2})+2\Delta (\mathbf{k},\Omega _{1},\Omega _{2})%
\right]
\end{equation*}%
in $\mathbf{k}$-space, and thus generates also $\Delta $. Consequently, we
will expand any angle-dependent correlation function $F$ as%
\begin{equation}
F(1,2)=F_{0}(r)+F_{\Delta }(r)\Delta (1,2)+F_{D}(r)D(1,2)+\cdots \ ,
\label{oz7}
\end{equation}%
neglecting all higher order terms. In other words, we assume that all
angular series expansions can be \textit{truncated} after these first three
terms, \textit{linear} with respect to the angular basis functions. Using
this spirit in the PY approximation, given by Eq. (\ref{w3b}), we obtain the
following PY correlation functions with \textit{orientational linearization}
(OL):%
\begin{eqnarray}
c^{\mathrm{PY-OL}}(1,2) &=&c_{0}(r)+c_{\Delta }(r)\Delta (1,2)+c_{D}(r)D(1,2)
\notag \\
&=&c_{0}(r)+\left( \alpha t\right) \ \left[ \ c_{\Delta }^{\dagger
}(r)\Delta (1,2)+c_{D}^{\dagger }(r)D(1,2)\ \right] ,  \label{oz10aa}
\end{eqnarray}%
and

\begin{eqnarray}
h^{\mathrm{PY-OL}}(1,2) &=&h_{0}(r)+h_{\Delta }(r)\Delta (1,2)+h_{D}(r)D(1,2)
\notag \\
&=&h_{0}(r)+\left( \alpha t\right) \ \left[ \ h_{\Delta }^{\dagger
}(r)\Delta (1,2)+h_{D}^{\dagger }(r)D(1,2)\ \right] .  \label{oz10bb}
\end{eqnarray}%
where%
\begin{equation}
\left. 
\begin{array}{c}
c_{0}(r)=\Lambda _{0}\text{\ }\sigma \delta \left( r-\sigma \right) \\ 
c_{\Delta }(r)=\Lambda _{\Delta }\text{\ }\sigma \delta \left( r-\sigma
\right) \\ 
c_{D}(r)=\Lambda _{D}\ \sigma \delta \left( r-\sigma \right)%
\end{array}%
\right\} \qquad \text{for \ }r\geq \sigma ,  \label{closure}
\end{equation}%
with%
\begin{equation}
\text{\ \ }\left\{ 
\begin{array}{l}
\Lambda _{0}=y_{0}^{\mathrm{PY}}(\sigma )\ t\text{\ \ \ \ \ \ \ \ \ \ \ \ \
\ \ \ \ \ } \\ 
\Lambda _{\Delta }=y_{\Delta }^{\mathrm{PY}}(\sigma )\ t\text{\ \ \ \ \ \ \
\ \ \ \ \ \ \ \ \ \ \ } \\ 
\Lambda _{D}=\left[ \ y_{D}^{\mathrm{PY}}(\sigma )+\alpha \ y_{0}^{\mathrm{PY%
}}(\sigma )\right] \ t = y_{D}^{\mathrm{PY}}(\sigma )\ t+\alpha \ \Lambda
_{0} ,%
\end{array}%
\right.  \label{oz21}
\end{equation}%
\begin{equation}
\left\{ 
\begin{array}{c}
y_{0}^{\mathrm{PY}}\left( r\right) =1+\gamma _{0}(r)\text{\ \ \ \ \ \ \ \ \
\ \ \ \ \ \ \ \ } \\ 
y_{\Delta }^{\mathrm{PY}}(r)=\gamma _{\Delta }(r)=\left( \alpha t\right) \
y_{\Delta }^{\dagger }(r)\text{ \ \ } \\ 
y_{D}^{\mathrm{PY}}(r)=\gamma _{D}(r)=\left( \alpha t\right) \
y_{D}^{\dagger }(r).\text{ \ }%
\end{array}%
\right.  \label{ysigma}
\end{equation}

Clearly for $f(1,2)$ no truncation is required, as the expansion 
\begin{equation}
\left\{ 
\begin{array}{c}
f_{0}(r)=f^{\mathrm{isoSHS}}(r)=f^{\mathrm{HS}}(r)+t\text{\ }\sigma \delta
\left( r-\sigma \right) \text{\ \ \ \ \ \ \ \ \ } \\ 
f_{\Delta }(r)=0\text{ \ \ \ \ \ \ \ \ \ \ \ \ \ \ \ \ \ \ \ \ \ \ \ \ \ \ \
\ \ \ \ \ \ \ \ \ \ \ \ \ \ \ \ \ \ \ \ \ \ \ \ } \\ 
f_{D}(r)=\left( \alpha t\right) \ \sigma \delta \left( r-\sigma \right) 
\text{ \ \ \ \ \ \ \ \ \ \ \ \ \ \ \ \ \ \ \ \ \ \ \ \ \ \ \ \ \ \ \ \ }%
\end{array}%
\right.
\end{equation}%
is exact.

It can be shown that $c(1,2)$ and $h(1,2)$ must have the same approximate
form in view of the OZ equation, Eq. (\ref{oz4}).

The solution of the original OZ equation (\ref{oz4}) is then equivalent to
the calculation of the radial coefficients $c_{0}(r),c_{\Delta }(r),c_{D}(r)$
and $h_{0}(r),h_{\Delta }(r),h_{D}(r)$, which are the projections of $c(1,2)$
and $h(1,2)$ onto the angular basis $\left\{ 1,\Delta ,D\right\} $. The 
\textit{core condition} on $h$, Eq. (\ref{oz6a}), becomes%
\begin{equation}
\left. 
\begin{array}{c}
h_{0}(r)=-1 \\ 
h_{\Delta }(r)=0 \\ 
h_{D}(r)=0%
\end{array}%
\right\} \text{ \ \ \ \ \ for \ }0<r<\sigma .  \label{oz10a}
\end{equation}

Note that in the zero density limit $\gamma (1,2)=\rho \int d\mathbf{r}_{3}\
\left\langle \ c(1,3)\ h(3,2)\ \right\rangle _{\Omega _{3}}$ must vanish,
and thus $y^{\mathrm{PY}}(1,2)\rightarrow 1$, i.e.

\begin{equation*}
\lim_{\rho \rightarrow 0}y_{0}^{\mathrm{PY}}(r)=1\text{, \ \ \ \ \ \ \ }%
\lim_{\rho \rightarrow 0}y_{\Delta }^{\mathrm{PY}}(r)=\lim_{\rho \rightarrow
0}y_{D}^{\mathrm{PY}}(r)=0,
\end{equation*}%
while both $c\left( 1,2\right) $ and $h\left( 1,2\right) $ must reduce to $%
f(1,2)$: 
\begin{equation*}
\begin{array}{c}
\lim_{\rho \rightarrow 0}F_{0}(r)=f_{0}(r)\text{ } \\ 
\text{\ }\lim_{\rho \rightarrow 0}F_{\Delta }(r)=0\text{ \ \ \ \ \ \ \ } \\ 
\lim_{\rho \rightarrow 0}F_{D}(r)=f_{D}(r)\text{ }%
\end{array}%
\text{\ \ \ \ \ \ }(F=c,h),
\end{equation*}%
and%
\begin{equation}
\lim_{\rho \rightarrow 0}\Lambda _{0}=t\text{, \ \ \ \ \ \ \ }\lim_{\rho
\rightarrow 0}\Lambda _{\Delta }=0\text{, \ \ \ \ \ \ }\lim_{\rho
\rightarrow 0}\Lambda _{D}=\alpha t\text{.}
\end{equation}

Moreover as $\alpha \rightarrow 0$ all $\Delta $- and $D$-coefficients of $%
c,h$ and $y$ vanish, so that the isotropic adhesion case is recovered.
Finally, it is also worth stressing that the \textit{same} $\delta $-term
arises in $c,$ $h$ and $g$, that is

\begin{equation*}
F(1,2)=F_{\mathrm{reg}}(1,2)+F_{\text{sing}}(1,2)\text{ \ \ \ \ \ \ (}%
F=c,h,g),
\end{equation*}%
where $F_{\mathrm{reg}}$ is the `regular' part (i.e., the part with no $%
\delta $-singularity, and -- at most -- some step discontinuities), while $%
F_{\text{sing}}(1,2)=\ \sigma \delta \left( r-\sigma \right) \Lambda (1,2)$
is the singular term representing the anisotropic surface adhesion ( with$\
\Lambda (1,2)=\Lambda _{0}+\Lambda _{\Delta }\Delta (1,2)+\Lambda _{D}D(1,2)$
\ ).

\subsection{Integral equations for the projections of $c$ and $h$}

In the following, we extend Wertheim theory \cite{Wertheim71} to our model,
in order to obtain the radial projections of $c$ and $h$.

\ \ The PY-OL approximation to the excess anisotropic part of the
correlation functions is 
\begin{equation}
\begin{array}{c}
c_{\mathrm{ex}}^{\mathrm{PY-OL}}(1,2)=c_{\Delta }(r)\Delta
(1,2)+c_{D}(r)D(1,2)\text{ } \\ 
h_{\mathrm{ex}}^{\mathrm{PY-OL}}(1,2)=h_{\Delta }(r)\Delta
(1,2)+h_{D}(r)D(1,2),%
\end{array}%
\end{equation}%
thus verifying the required property $\left\langle \ c_{\mathrm{ex}}(1,3)\
\right\rangle _{\Omega _{3}}=\left\langle \ h_{\mathrm{ex}}(3,2)\
\right\rangle _{\Omega _{3}}=0$ described in Section III, and allowing the
splitting of the molecular OZ equation into a reference and an excess part.

The first part is the \textit{reference} PY equation, and coincides with
that solved by Baxter for the fluid with isotropic adhesion \cite%
{Baxter68,Baxter71}:

\begin{equation}
\left\{ 
\begin{array}{cc}
h_{0}(r)=c_{0}(r)+\rho \ (h_{0}\star c_{0})\text{ \ \ \ \ \ \ \ \ \ \ \ \ \
\ \ } &  \\ 
h_{0}(r)=-1\text{\ \ \ \ \ \ \ \ \ \ \ \ \ \ \ \ \ \ \ \ \ \ \ \ \ \ \ \ \ \
\ \ \ \ \ } & 0<r<\sigma \\ 
c_{0}(r)=\Lambda _{0}\ \sigma \delta \left( r-\sigma \right) \text{ \ \ \ \
\ \ \ \ \ \ \ \ \ \ \ \ \ \ \ \ } & r\geq \sigma ,%
\end{array}%
\right.  \label{w1}
\end{equation}%
where the symbol $\star $ denotes spatial convolution, i.e. $(A\star
B)(r_{12})=\int A(r_{13})B(r_{32})\;d\mathbf{r}_{3}$.

The second part is the \textit{excess PY-OL equation}, given by Eq. (\ref%
{ozeq2}) coupled with the PY-OL closure. Following an extension of
Wertheim's approach, as described in detail in Appendix A, Eq. (\ref{ozeq2})
can be splitted into the following system for the $\Delta -$ and $D-$%
projections of $c$ and $h:$ 
\begin{equation}
\left\{ 
\begin{array}{c}
h_{\Delta }(r)=c_{\Delta }(r)+\frac{1}{3}\rho ~\left( \ c_{\Delta }\star
h_{\Delta }+2\ c_{D}^{0}\star h_{D}^{0}\ \right) \text{\ \ \ \ \ \ \ \ \ \ \ 
} \\ 
h_{D}^{0}(r)=c_{D}^{0}(r)+\frac{1}{3}\rho ~\left( \ c_{\Delta }\star
h_{D}^{0}+c_{D}^{0}\star h_{\Delta }+c_{D}^{0}\star h_{D}^{0}\ \right) ,%
\end{array}%
\right.
\end{equation}%
where $c_{D}^{0}(r)$ and $h_{D}^{0}(r)$ are defined by the relationship%
\begin{equation}
F_{D}^{0}(r)=F_{D}(r)-3\int_{r}^{\infty }\frac{F_{D}(x)}{x}\ dx\text{ \ \ \
\ \ \ \ \ }(F=c,h),  \label{oz12}
\end{equation}%
whose inverse is \cite{Wertheim71} 
\begin{equation}
F_{D}(r)=F_{D}^{0}(r)-\frac{3}{r^{3}}\int_{0}^{r}F_{D}^{0}(x)\ x^{2}\ dx.
\label{oz12a}
\end{equation}

The core conditions become%
\begin{equation}
\left. 
\begin{array}{c}
\text{ \ }h_{\Delta }(r)=0\text{ \ \ \ \ \ \ } \\ 
h_{D}^{0}(r)=-3K%
\end{array}%
\right\} \text{ \ \ \ \ \ for \ }0<r<\sigma ,  \label{oz14a}
\end{equation}%
with 
\begin{equation}
K=\int_{\sigma ^{-}}^{\infty }\frac{h_{D}(x)}{x}\ dx=K_{\mathrm{reg}%
}+\Lambda _{D}\ ,  \label{oz14b}
\end{equation}%
\begin{equation}
K_{\mathrm{reg}}=\int_{\sigma }^{\infty }\frac{h_{D,\text{\textrm{reg}}}(r)}{%
r}\ dr.  \label{oz14bb}
\end{equation}%
Note that the presence of the $\delta $-singularity in $h_{D}(x)$ requires
the specification of $\sigma ^{-}$ as lower integration limit, unlike the
case of Ref. I where only the regular part $K_{\mathrm{reg}}$ is present.
Moreover, since $h_{D}(r)=\alpha t\ \ h_{D}^{\dagger }(r)$, from Eq. (\ref%
{oz14b}) one could also write%
\begin{equation}
K=\alpha t\ \mathcal{K}\text{ ,}
\end{equation}%
which shows that $K$ is related to the anisotropy degree, and vanishes both
in the symmetric adhesion case ($\alpha =0$) and in the HS limit ($t=0$).
Since $h_{D}(r)\rightarrow f_{D}(r)=\left( \alpha t\right) \sigma \delta
(r-\sigma )$ in the zero density limit, one then finds that

\begin{equation}
\lim_{\eta \rightarrow 0}K=\alpha t.
\end{equation}

Finally, the PY-OL closure for the new DCFs reads%
\begin{equation}
\left. 
\begin{array}{c}
c_{\Delta }(r)=\Lambda _{\Delta }\ \sigma \delta \left( r-\sigma \right) \\ 
c_{D}^{0}(r)=\Lambda _{D}\ \sigma \delta \left( r-\sigma \right)%
\end{array}%
\right\} \text{ \ \ \ \ \ }r\geq \sigma  \label{oz14c}
\end{equation}%
(for simplicity, here and in the following we omit the superscript PY-OL).

\subsection{Decoupling of the integral equations}

It is possible to \textit{decouple} the two equations for $\Delta $- and $D$%
-coefficients, by introducing two new unknown functions, which are linear
combinations of the previous ones. As shown in Appendix A, if we define $%
F_{1}(r)$ and $F_{2}(r)$ $\left( F=c,h\right) $ through the relations

\begin{equation*}
\left\{ 
\begin{array}{c}
F_{1}\left( r\right) =\left( 3\mathcal{L}_{1}\right) ^{-1}\left[ F_{\Delta
}(r)-F_{D}^{0}(r)\right] \text{ } \\ 
F_{2}\left( r\right) =\left( 3\mathcal{L}_{2}\right) ^{-1}\left[ F_{\Delta
}(r)+2F_{D}^{0}(r)\right]%
\end{array}%
\right. \text{\ \ \ \ }\left( F=c,h\right) ,
\end{equation*}%
then we get the OZ equations

\begin{equation*}
\left\{ 
\begin{array}{c}
h_{1}(r)=c_{1}(r)+\rho _{1}\ (h_{1}\star c_{1})\text{ } \\ 
h_{2}(r)=c_{2}(r)+\rho _{2}\ (h_{2}\star c_{2}),%
\end{array}%
\right.
\end{equation*}%
with the following densities and core conditions

\begin{equation*}
\left\{ 
\begin{array}{c}
\rho _{1}=\mathcal{L}_{1}\ \rho \text{ } \\ 
\rho _{2}=\mathcal{L}_{2}\ \rho ,%
\end{array}%
\right. \qquad \qquad \left\{ 
\begin{array}{c}
h_{1}(r)=K/\mathcal{L}_{1}\text{ \ \ \ } \\ 
h_{2}(r)=-2K/\mathcal{L}_{2}%
\end{array}%
\right. \text{ \ \ \ \ \ for \ }0<r<\sigma .
\end{equation*}%
\ \ 

The decoupling of the three different projections of $c$ and $h$ is
remarkable: the molecular anisotropic OZ equation reduces to a set of three
radial integral relations, which may be regarded as OZ equations for three
`hypothetical' fluids (labelled as $0,1,2$) with \textit{spherically
symmetric} interactions. We stress that there is not a unique solution to
the decoupling problem, since - in principle - there exist infinite possible
choices for $\left( \mathcal{L}_{1},\mathcal{L}_{2}\right) $. The final
results are clearly independent of the values of $\left( \mathcal{L}_{1},%
\mathcal{L}_{2}\right) $.

In the present paper, we adopt Wertheim's choice, i.e. $\mathcal{L}_{1}=-K$
and $\mathcal{L}_{2}=2K$, which leads to%
\begin{equation}
\left\{ 
\begin{array}{c}
F_{1}\left( r\right) =\frac{1}{3K}\left[ F_{D}^{0}(r)-F_{\Delta }(r)\right] 
\text{ } \\ 
F_{2}\left( r\right) =\frac{1}{3K}\left[ F_{D}^{0}(r)+\frac{1}{2}F_{\Delta
}(r)\right]%
\end{array}%
\right. \text{\ \ \ \ }\left( F=c,h\right) ,
\end{equation}

\begin{equation}
\left\{ 
\begin{array}{c}
\rho _{1}=-K\rho \text{ } \\ 
\rho _{2}=2K\rho ,%
\end{array}%
\right. \qquad \qquad \left\{ 
\begin{array}{c}
h_{1}(r)=-1 \\ 
h_{2}(r)=-1%
\end{array}%
\right. \text{ \ \ \ \ \ for \ }0<r<\sigma
\end{equation}%
(in Ref. I, $F_{1}$ and $F_{2}$ were denoted as $F_{-}$ and $F_{+}$,
respectively).

Note that the auxiliary fluids have densities different from that of the
reference fluid (the negative sign of $\rho _{1}$ poses no special
difficulty).

We can also write

\begin{equation}
F_{m}\left( r\right) =F_{m,\mathrm{reg}}(r)+\Lambda _{m}\ \sigma \delta
\left( r-\sigma \right) ,  \label{oz15a}
\end{equation}%
with

\begin{equation}
\left\{ 
\begin{array}{c}
F_{1,\mathrm{reg}}\left( r\right) =\frac{1}{3K}\left[ F_{D,\mathrm{reg}%
}^{0}(r)-F_{\Delta ,\mathrm{reg}}(r)\right] \text{ } \\ 
F_{2,\mathrm{reg}}\left( r\right) =\frac{1}{3K}\left[ F_{D,\mathrm{reg}%
}^{0}(r)+\frac{1}{2}F_{\Delta ,\mathrm{reg}}(r)\right]%
\end{array}%
\right.  \label{oz15b}
\end{equation}%
and%
\begin{equation}
\left\{ 
\begin{array}{c}
\Lambda _{1}=\frac{1}{3K}\left( \Lambda _{D}-\Lambda _{\Delta }\text{\ }%
\right) \text{ \ \ \ \ \ \ \ \ \ \ \ \ \ \ \ \ \ \ \ \ \ \ \ \ \ \ \ \ \ \ \
\ \ \ \ \ \ \ \ } \\ 
=\frac{1}{3K}\left[ h_{D,\text{\textrm{reg}}}(\sigma ^{+})-h_{\Delta ,\text{%
\textrm{reg}}}(\sigma ^{+})\right] \ t\text{ }+\alpha \ \frac{1}{3K}\
\Lambda _{0}\text{ \ } \\ 
\text{\ \ \ \ \ \ } \\ 
\Lambda _{2}=\frac{1}{3K}\left( \Lambda _{D}+\frac{1}{2}\Lambda _{\Delta
}\right) \text{ \ \ \ \ \ \ \ \ \ \ \ \ \ \ \ \ \ \ \ \ \ \ \ \ \ \ \ \ \ \
\ \ \ \ \ \ \ \ } \\ 
\text{ \ \ }=\frac{1}{3K}\left[ h_{D,\text{\textrm{reg}}}(\sigma ^{+})+\frac{%
1}{2}h_{\Delta ,\text{\textrm{reg}}}(\sigma ^{+})\right] \ t\text{ }+\alpha
\ \frac{1}{3K}\ \Lambda _{0}\text{.\ \ }%
\end{array}%
\right.  \label{oz16}
\end{equation}%
(since $\gamma _{\ldots }(\sigma )=h_{\ldots ,\text{\textrm{reg}}}(\sigma
^{+})-c_{\ldots ,\text{\textrm{reg}}}(\sigma ^{+})$, and $c_{\ldots ,\text{%
\textrm{reg}}}(\sigma ^{+})=0$ within the PY-OL closure).

Knowing the correlation functions $F_{1}(r)$ and $F_{2}(r)$ $\left( \text{%
with }F=c,h\right) $, one can derive $F_{\Delta }\left( r\right)
,F_{D}^{0}\left( r\right) $, i.e. 
\begin{equation}
\left\{ 
\begin{array}{c}
F_{\Delta }\left( r\right) =2K\left[ F_{2}\left( r\right) -F_{1}\left(
r\right) \right] \ \text{\ \ \ \ } \\ 
F_{D}^{0}\left( r\right) =2K\left[ F_{2}\left( r\right) +\frac{1}{2}%
F_{1}\left( r\right) \right] \ ,%
\end{array}%
\right.  \label{so11}
\end{equation}%
and 
\begin{equation}
\left\{ 
\begin{array}{c}
\Lambda _{\Delta }=2K\left( \Lambda _{2}-\Lambda _{1}\right) \text{\ \ \ \ }
\\ 
\Lambda _{D}=K\left( 2\Lambda _{2}+\Lambda _{1}\right) \text{ .\ \ }%
\end{array}%
\right.  \label{c10}
\end{equation}

Finally, from $F_{\Delta }\left( r\right) ,F_{D}^{0}\left( r\right) $ one
has to evaluate $F_{\Delta }\left( r\right) ,F_{D}\left( r\right) $, by
employing Eq. (\ref{oz12a}). We note the following points:

i) Insertion of $h_{D}^{0}(r)=h_{D,\text{\textrm{reg}}}^{0}(r)+\Lambda _{D}\
\sigma \delta (r-\sigma )$ into Eq. (\ref{oz12a}) yields $h_{D}(r)=h_{D,%
\text{\textrm{reg}}}(r)+\Lambda _{D}\ \sigma \delta (r-\sigma ),$ with%
\begin{equation}
h_{D,\text{\textrm{reg}}}(r)=\left\{ 
\begin{array}{cc}
0\text{ \ \ \ \ \ \ \ \ \ \ \ \ \ \ \ \ \ \ \ \ \ \ \ \ \ \ \ \ \ \ \ \ \ \
\ \ \ \ \ \ \ \ \ \ \ }0<r<\sigma ,\text{\ \ \ \ \ \ } &  \\ 
h_{D,\text{\textrm{reg}}}^{0}(r)+3r^{-3}\left[ K_{\mathrm{reg}}\sigma
^{3}-\int_{\sigma }^{r}h_{D,\text{\textrm{reg}}}^{0}(x)\ x^{2}\ dx\right] 
\text{ \ \ \ \ \ \ \ } & r>\sigma .%
\end{array}%
\right.  \label{c11}
\end{equation}%
At $r=2\sigma $ $h_{D,\text{\textrm{reg}}}$ and $h_{D,\text{\textrm{reg}}%
}^{0}$ have the same discontinuity. We also get

\begin{equation}
h_{D,\text{\textrm{reg}}}(\sigma ^{+})=h_{D,\text{\textrm{reg}}}^{0}(\sigma
^{+})+3K_{\mathrm{reg}}.  \label{f7}
\end{equation}%
Clearly, these results must agree with those obtained from Eq. (\ref{oz12}),
i.e. 
\begin{equation*}
\begin{array}{c}
h_{D}^{0}(r)=h_{D}(r)-3\psi (r),\text{ \ \ \ \ \ \ \ \ \ \ \ \ \ \ \ \ \ \ \
\ \ \ \ \ \ \ \ \ \ \ \ \ \ \ \ \ \ \ \ \ \ \ \ \ \ \ \ \ } \\ 
\psi (r)\equiv \int_{r}^{\infty }h_{D}(x)\ x^{-1}\ dx=\Lambda _{D}\ \theta
\left( \sigma -r\right) +\int_{r}^{\infty }h_{D,\text{\textrm{reg}}}(x)\
x^{-1}\ dx.%
\end{array}%
\text{\ }
\end{equation*}%
In order to recover Eq. (\ref{f7}) along this second route, note that $\psi
(r)$ is not continuous at $r=\sigma $. In fact, from Eqs. (\ref{oz14b}) and (%
\ref{oz14bb}) follows $\psi (\sigma ^{-})=K$ whereas $\psi (\sigma ^{+})=K_{%
\mathrm{reg}}.$

ii) Similarly, for $c_{D}(r)$ we obtain $c_{D}(r)=c_{D,\text{\textrm{reg}}%
}(r)+\Lambda _{D}\ \sigma \delta (r-\sigma )$, with

\begin{equation}
c_{D,\text{\textrm{reg}}}(r)=c_{D,\text{\textrm{reg}}}^{0}(r)-3r^{-3}\left[
\int_{0}^{r}\ c_{D,\text{\textrm{reg}}}^{0}(x)x^{2}\ dx+\Lambda _{D}\sigma
^{3}\ \theta (r-\sigma )\right] ,  \label{c12}
\end{equation}%
since $\int_{0}^{r}\ \delta (x-\sigma )x^{2}dx=\sigma ^{2}\theta (r-\sigma )$%
. On the other hand, from Eq. (\ref{oz12}) one easily finds that%
\begin{equation}  \label{cDcD0}
c_{D}(r)=c_{D}^{0}(r)\text{ \ \ \ for \ }r\geq \sigma .
\end{equation}%
\ \ \ 

iii) By applying the relationship (\ref{oz12a}) to $c_{D}(r)$, using Eq. (%
\ref{cDcD0}) and noticing that $c_{D}(x)=0$ for $r>\sigma $ within the PY-OL
approximation, leads to a \textit{sum rule}: 
\begin{equation}
\int_{0}^{\infty }c_{D}^{0}(x)\ x^{2}\ dx=\int_{0}^{\sigma }\ c_{D,\text{%
\textrm{reg}}}^{0}(x)x^{2}\ dx+\Lambda _{D}\sigma ^{3}\ =0,  \label{f4b}
\end{equation}%
that we will exploit later.

\section{ANALYTIC\ SOLUTION}

We have seen that the molecular PY-OL integral equation (IE) for our \textit{%
anisotropic}-SHS model splits into three IE's 
\begin{equation}
\left\{ 
\begin{array}{cc}
h_{m}(r)=c_{m}(r)+\rho _{m}\ (h_{m}\star c_{m})\text{\ \ } &  \\ 
h_{m}(r)=-1\text{\ \ \ \ \ \ \ \ \ \ \ \ \ \ \ \ \ \ \ \ \ \ \ \ \ \ } & 
0<r<\sigma \\ 
c_{m}(r)=\Lambda _{m}\ \sigma \delta \left( r-\sigma \right) \text{ \ \ \ \
\ \ \ \ \ \ } & r\geq \sigma%
\end{array}%
\right. \qquad \left( m=0,1,2\right) ,  \label{gpy1}
\end{equation}%
where

\begin{equation}
\left\{ 
\begin{array}{c}
\rho _{0}=\rho \text{ \ \ \ \ \ } \\ 
\rho _{1}=-K\rho \text{ } \\ 
\rho _{2}=2K\rho ,%
\end{array}%
\right.  \label{gpy2}
\end{equation}%
and the `amplitudes' of the adhesive $\delta $-terms are 
\begin{equation}
\left\{ 
\begin{array}{c}
\text{ }\Lambda _{0}=\left[ 1+h_{0,\text{\textrm{reg}}}(\sigma ^{+})\right]
\ t=y_{0}^{\mathrm{PY}}(\sigma )t\text{ }\text{ \ \ \ \ \ \ \ \ \ \ \ \ \ \
\ \ \ \ \ \ \ \ \ \ \ \ \ \ \ \ } \\ 
\text{ }\Lambda _{m}=h_{m,\text{\textrm{reg}}}(\sigma ^{+})\ t\text{ }+%
\mathcal{P} = y_{m}^{\mathrm{PY}}(\sigma )t+\mathcal{P}\text{ \ \ \ \ \ }%
\left( m=1,2\right) ,\text{ }%
\end{array}%
\right.  \label{gpy3}
\end{equation}%
with%
\begin{equation}
\mathcal{P}=\frac{1}{3}\ \frac{\alpha t\ y_{0}^{\mathrm{PY}}(\sigma )\ }{K}+%
\frac{K_{\mathrm{reg}}}{K}\ t=\frac{1}{3}\ \frac{\alpha \ \Lambda _{0}\ }{K}+%
\frac{K_{\mathrm{reg}}}{K}\ t.  \label{gpy4}
\end{equation}%
Here, the new expressions of $\Lambda _{1}$ and $\Lambda _{2}$ have been
obtained from Eqs. (\ref{oz16}) with the help of Eqs. (\ref{f7}) and (\ref%
{oz15b}).

The essential difference with respect to Ref. I lies in the closure, which
is -- of course -- related to the model potential. While Wertheim's paper on
DHS \cite{Wertheim71} employed the MSA closure, which performs properly for
long-ranged electrostatic potentials at low strength of interaction, our
PY-OL closure is more appropriate for the short-ranged potential of the
present model.

The first integral equation IE0 is fully independent, whereas IE1 and IE2
depend on the solution of IE0 (unlike the case of Ref. I), because of the
presence of $\Lambda _{0}$ inside $\Lambda _{1\text{ }}$and $\Lambda _{2}$.
While IE0 is exactly the PY equation for the \textit{reference} SHS with 
\textit{isotropic} adhesion solved by Baxter \cite{Baxter68,Baxter71}, IE1
and IE2 are different from both Wertheim's MSA solution for DHS and Baxter's
PY solution for SHS. We remark that the closures for IE1 and IE2 are not PY
as $\Lambda _{1\text{ }}$and $\Lambda _{2}$ -- given by Eq. (\ref{gpy3}) --
differ, by the term $\mathcal{P}$, from those appropriate for the PY choice,
corresponding to $\Lambda _{m}^{\mathrm{PY}}=y_{m}^{\mathrm{PY}}(\sigma )t$.

Consequently, IE1 and IE2 can be rekoned as belonging to a class of \textit{%
generalized PY }\ (GPY) \textit{approximations}, introduced in Ref. [39],
which admit an analytic solution. Thus, the PY-OL closure for $c(1,2)$ leads
to a PY integral equation for $c_{0}(r)$, coupled to a two GPY integral
equations for $c_{1}(r)$ and $c_{2}(r)$ (which are linear combinations of $%
c_{\Delta }(r)$ and $c_{D}^{0}(r)$).

On comparing the three IE's and their closures given by Eq. (\ref{gpy1}), it
is apparent that they have exactly the same form, but differ by the density $%
\rho _{m}$ and the expression for $\Lambda _{m}$. The first integral
equation IE0 corresponds to an isotropic SHS fluid with density $\rho $. On
the other hand, IE1 and IE2 refer to `auxiliary' isotropic SHS fluids with
densities $\rho _{1}$ and $\rho _{2}$, and adhesion parameters $\Lambda _{1}$
and $\Lambda _{2}$, respectively. Note that, according to Eqs. (\ref{oz16})
, $\Lambda _{m}$ is not evaluated at the actual density $\rho _{m}$ of the
auxiliary fluid, but at the real density $\rho $ . These remarks strongly
suggests that the solutions of IE0, IE1 and IE2 can be expressed in terms of
an \textit{unique} solution -- the PY one for isotropic SHS -- by changing
only $\rho _{m}$ and $\Lambda _{m}$. This can be achieved by the formal
mapping 
\begin{equation}
\left\{ 
\begin{array}{c}
F_{0}(r)=F^{\mathrm{isoSHS}}(r;\eta _{0},\Lambda _{0})\text{ } \\ 
F_{1}(r)=F^{\mathrm{isoSHS}}(r;\eta _{1},\Lambda _{1})\text{ } \\ 
F_{2}(r)=F^{\mathrm{isoSHS}}(r;\eta _{2},\Lambda _{2}),%
\end{array}%
\right. \qquad \quad (F=q,c,h)  \label{gpy5}
\end{equation}%
where $\eta _{0}=\eta $ is the \textit{real} volume fraction, while $\eta
_{1}$ and $\eta _{2}$ are `modified volume fractions' of the `auxiliary'
fluids $1$ and $2$, i.e.,\ 
\begin{equation}
\left\{ 
\begin{array}{c}
\eta _{0}=\eta \equiv \left( \pi /6\right) \rho \sigma ^{3}\text{ \ \ } \\ 
\eta _{1}=-K\eta \text{ \ \ \ \ \ \ \ \ \ \ \ \ \ } \\ 
\eta _{2}=2K\eta \text{\ .\ \ \ \ \ \ \ \ \ \ \ \ \ }%
\end{array}%
\right.  \label{gpy6}
\end{equation}%
In Eqs. (\ref{gpy5}) $q(r)$ denotes the Baxter factor correlation function,
introduced in the next Subsection.

It is worth noting that this result for SHS mirrors the analog of the MSA
solution for DHS \cite{Wertheim71} where all the three harmonic coefficients
of can be expressed similarly, in terms of a single PY solution for the
reference HS fluid.

\bigskip

\subsection{Baxter factorization}

We shall now solve Eqs. ($\ref{gpy1}$) by using the Wiener-Hopf
factorization due to Baxter \cite{Baxter71}. Let us recall its basic steps.
After Fourier transforming the OZ equation for a one-component fluid with
spherically symmetric interactions, one assumes the following factorization: 
\begin{equation}
\begin{array}{c}
1-\rho \widetilde{c}\left( k\right) =Q(k)Q(-k), \\ 
Q(k)=1-2\pi \rho \int_{0}^{\infty }q(r)\ e^{ikr}dr.%
\end{array}
\label{bf1}
\end{equation}%
Then it can be shown that the introduction of the `factor correlation
function' $q(r)$ allows the OZ equation to be cast into the form \cite%
{Baxter71}

\begin{equation}
\left\{ 
\begin{array}{c}
rc\left( r\right) =-q^{\prime }(r)+2\pi \rho \int_{0}^{\infty }du\ q\left(
u\right) q^{\prime }\left( r+u\right) ,\text{ \ \ \ \ \ \ \ \ \ \ } \\ 
rh\left( r\right) =-q^{\prime }(r)+2\pi \rho \int_{0}^{\infty }du\ q\left(
u\right) \left( r-u\right) h\left( |r-u|\right) ,%
\end{array}%
\right.  \label{ie2b}
\end{equation}%
where the prime denotes differentiation with respect to $r$. Solving these
Baxter equations is tantamount to determining -- within a chosen closure --
the function $q(r)$, from which $c\left( r\right) $ and $h\left( r\right) $
can be easily calculated. It is also necessary to remember that, for \textit{%
all} closures leading to $c\left( r\right) =0$ for $r>\sigma $, one finds $%
q\left( r\right) =0$ for $r>\sigma $ \cite{Gazzillo04}.

On applying Baxter's factorization to Eqs. ($\ref{gpy1}$), we get 
\begin{equation}
rh_{m}\left( r\right) =-q_{m}^{\prime }(r)+2\pi \rho _{m}\int_{0}^{\sigma
}du\ q_{m}\left( u\right) \left( r-u\right) h_{m}\left( |r-u|\right) .
\label{ie3}
\end{equation}%
with $m=0,1,2$. Now the closure $c_{m}(r)=\Lambda _{m}\ \sigma \delta \left(
r-\sigma \right) $ for \ $r\geq \sigma $ implies that the same $\delta $%
-term must appear in $h_{m}\left( r\right) $. Thus, for $0\leq r\leq \sigma $%
, using $h_{m}(r)=-1+\Lambda _{m}\ \sigma \delta \left( r-\sigma \right) $,
we find

\begin{equation*}
q_{m}^{\prime }(r)=a_{m}r+b_{m}\sigma -\Lambda _{m}\ \sigma ^{2}\delta
\left( r-\sigma \right) ,
\end{equation*}%
with%
\begin{equation}
\left\{ 
\begin{array}{c}
a_{m}=\ 1-2\pi \rho _{m}\int_{0}^{\sigma }du\ q_{m}\left( u\right) \ , \\ 
b_{m}\sigma =\ 2\pi \rho _{m}\int_{0}^{\sigma }du\ q_{m}\left( u\right) \ u\
.\text{ \ \ }%
\end{array}%
\right.  \label{ie4}
\end{equation}

The $\delta $-term of $q_{m}^{\prime }(r)$ means that $q_{m}(r)$ has a
discontinuity $q_{m}(\sigma ^{+})-q_{m}(\sigma ^{-})=-\Lambda _{m}\sigma
^{2} $, with $q_{m}(\sigma ^{+})=0.$ Integrating $q_{m}^{\prime }(r)$,
substituting this result into Eqs. ($\ref{ie4}$), and solving the
corresponding algebraic system, we find the following solution

\begin{equation}
q_{m}(r)=\left\{ 
\begin{array}{cc}
\ \frac{1}{2}a_{m}(r-\sigma )^{2}+\left( a_{m}+b_{m}\right) \sigma (r-\sigma
)\ +\Lambda _{m}\ \sigma ^{2} & \text{ \ \ \ \ }0\leq r\leq \sigma , \\ 
0 & \text{ \ \ \ otherwise,}%
\end{array}%
\right.  \label{so1}
\end{equation}%
\begin{eqnarray}
a_{m} &=&\ a^{\mathrm{HS}}(\eta _{m})-\frac{12\eta _{m}\ \Lambda _{m}\ }{%
1-\eta _{m}}\   \label{so2} \\
&&  \notag \\
b_{m} &=&\ b^{\mathrm{HS}}(\eta _{m})+\frac{6\eta _{m}\ \Lambda _{m}\ }{%
1-\eta _{m}}\ \   \label{so3} \\
&&  \notag \\
\text{\ }\eta _{m} &=&\left( \pi /6\right) \rho _{m}\sigma ^{3}\text{ \ \ \ }
\\
&&  \notag \\
a^{\mathrm{HS}}(x) &=&\frac{1+2x}{\left( 1-x\right) ^{2}},\text{ \ \ \ \ \ \
\ }b^{\mathrm{HS}}(x)=-\frac{3x}{2\left( 1-x\right) ^{2}},
\end{eqnarray}

From the first of Eqs. ($\ref{ie2b}$) we get the DCFs $c_{m}(r)=c_{m,\text{%
\textrm{reg}}}(r)+\ \Lambda _{m}\ \sigma \delta (r-\sigma )$, where $c_{m,%
\text{\textrm{reg}}}(r)=0$ for $r\geq \sigma ,$ and for $0<r<\sigma $

\begin{eqnarray}
c_{m,\text{\textrm{reg}}}(r) &=&-\frac{1}{2}\eta _{m}\ a_{m}^{2}\left( \frac{%
r}{\sigma }\right) ^{3}+6\eta _{m}\ \left[ \ (a_{m}+b_{m})^{2}-2a_{m}\
\Lambda _{m}\ \right] \left( \frac{r}{\sigma }\right)  \notag \\
&&-a_{m}^{2}-12\eta _{m}\ \Lambda _{m}^{2}\left( \frac{r}{\sigma }\right)
^{-1}.
\end{eqnarray}

The second of Eqs. ($\ref{ie2b}$) yields the total correlation functions $%
h_{m}(r)=h_{m,\text{\textrm{reg}}}(r)+\ \Lambda _{m}\ \sigma \delta
(r-\sigma ).$ For $r>\sigma $, Eqs. ($\ref{ie3}$) becomes%
\begin{equation}
H_{m,\text{\textrm{reg}}}\left( r\right) =12\eta _{m}\ \sigma ^{-3}\left\{ 
\begin{array}{cc}
\begin{array}{c}
\int_{0}^{r-\sigma }du\ q_{m}\left( u\right) \ H_{m,\text{\textrm{reg}}%
}\left( r-u\right) \text{ \ \ \ \ \ \ \ \ \ \ \ \ \ \ \ \ \ \ \ \ \ \ \ \ \
\ \ \ \ \ \ } \\ 
+\ \int_{r-\sigma }^{\sigma }du\ q_{m}\left( u\right) \left( u-r\right)
+\Lambda _{m}\sigma ^{2}\ q_{m}\left( r-\sigma \right) \text{ \ \ \ }%
\end{array}
& \sigma <r<2\sigma , \\ 
&  \\ 
\int_{0}^{\sigma }du\ q_{m}\left( u\right) \ H_{m,\text{\textrm{reg}}}\left(
r-u\right) \text{ \ \ \ \ \ \ \ \ \ \ \ \ \ }r>2\sigma ,\text{\ \ \ \ \ \ \
\ \ \ \ } & 
\end{array}%
\right.  \label{so7}
\end{equation}%
where $H_{m}\left( r\right) \equiv rh_{m}\left( r\right) $. Due to the last
term of Eq. ($\ref{so7}$) and the discontinuity of $q_{m}\left( r\right) $
at $r=\sigma $, $h_{m,\text{\textrm{reg}}}(r)$ has a jump of at $r=2\sigma $ 
\cite{Kranendonk88,Miller04}: $h_{m,\text{\textrm{reg}}}(2\sigma ^{+})-h_{m,%
\text{\textrm{reg}}}(2\sigma ^{-})=-6\eta _{m}\ \Lambda _{m}^{2}.$

\bigskip

\subsection{An important relationship}

In Appendix B it is shown that a remarkable consequence of the sum rule (\ref%
{f4b}) is the condition%
\begin{equation}
a_{2}=a_{1}\text{ ,}  \label{so10}
\end{equation}%
that will play a significant role in the determination of the unknown
parameters $\Lambda _{1},$ $\Lambda _{2}$ and $K$ \ (see Appendix B).

\bigskip

\subsection{Reference fluid coefficients}

The $m=0$ case corresponds to Baxter's PY results for the reference fluid of
isotropic SHS particles \cite{Baxter68,Baxter71}. We have: $q_{0}(r)=q^{%
\mathrm{isoSHS}}(r;\eta ,\Lambda _{0}),$ and%
\begin{equation}
\left\{ 
\begin{array}{c}
c_{0}(r)=c^{\mathrm{isoSHS}}(r;\eta ,\Lambda _{0})=c_{\text{\textrm{reg}}}^{%
\mathrm{isoSHS}}(r;\eta ,\Lambda _{0})+\Lambda _{0}\ \sigma \delta (r-\sigma
)\text{ } \\ 
h_{0}(r)=h^{\mathrm{isoSHS}}(r;\eta ,\Lambda _{0})=h_{\text{\textrm{reg}}}^{%
\mathrm{isoSHS}}(r;\eta ,\Lambda _{0})+\Lambda _{0}\ \sigma \delta (r-\sigma
)%
\end{array}%
\right.  \label{f1}
\end{equation}%
(for simplicity, we omit -- here and in the following -- the superscript PY).

\subsection{$\Delta -$ and $D-$coefficients}

We write $q_{m}(r)=q^{\mathrm{isoSHS}}(r;\eta _{m},\Lambda _{m})$ with $%
m=1,2 $. Then,

i) For the $\Delta $\textit{-coefficients}, after recalling Eq. (\ref{oz15a}%
) and exploiting Eqs. (\ref{c10}), we end up with:%
\begin{equation}
\left\{ 
\begin{array}{c}
c_{\Delta }(r)=2K\left[ c_{0,\text{\textrm{reg}}}(r;2K\eta ,\Lambda
_{2})-c_{0,\text{\textrm{reg}}}(r;-K\eta ,\Lambda _{1})\right] +\Lambda
_{\Delta }\ \sigma \delta (r-\sigma )\text{\ \ \ } \\ 
h_{\Delta }(r)=2K\left[ h_{0,\text{\textrm{reg}}}(r;2K\eta ,\Lambda
_{2})-h_{0,\text{\textrm{reg}}}(r;-K\eta ,\Lambda _{1})\right] +\Lambda
_{\Delta }\ \sigma \delta (r-\sigma ).\text{ \ }%
\end{array}%
\right.  \label{f2}
\end{equation}

ii) For the $D$\textit{-coefficients}, we get

\begin{equation}
\left\{ 
\begin{array}{c}
c_{D}^{0}(r)=2K\left[ c_{0,\text{\textrm{reg}}}(r;2K\eta ,\Lambda _{2})+%
\frac{1}{2}c_{0,\text{\textrm{reg}}}(r;-K\eta ,\Lambda _{1})\right] \
+\Lambda _{D}\ \sigma \delta (r-\sigma )\text{ \ } \\ 
h_{D}^{0}(r)=2K\left[ h_{0,\text{\textrm{reg}}}(r;2K\eta ,\Lambda _{2})+%
\frac{1}{2}h_{0,\text{\textrm{reg}}}(r;-K\eta ,\Lambda _{1})\right] \
+\Lambda _{D}\ \sigma \delta (r-\sigma ).\text{ }%
\end{array}%
\right.  \label{f3}
\end{equation}%
Finally, from $c_{D}^{0}(r)$ and $h_{D}^{0}(r)$ we can calculate $c_{D}(r)$
and $h_{D}(r)$, as described by Eqs. (\ref{c12}) and (\ref{c11}),
respectively.

In short, a) our PY-OL solution -- $\left\{ c_{0},c_{\Delta },c_{D}\right\} $
and $\left\{ h_{0},h_{\Delta },h_{D}\right\} $ -- satisfies both the PY
closures and the core conditions; b) all\textit{\ }coefficients contain a
surface adhesive $\delta -$term; c) $\left\{ h_{0},h_{\Delta },h_{D}\right\} 
$ all exhibit a step discontinuity at $r=2\sigma $.

\bigskip

\section{EVALUATION OF THE PARAMETERS $K$, $\Lambda _{1}$ AND $\Lambda _{2}$}

The calculation of the Baxter functions $q_{m}s$ ($m=0,1,2$) requires the
evaluation of $K,$ $\Lambda _{1}$, and $\Lambda _{2}$, for a given set of $%
\alpha ,\eta $ and $t$ values, a task that we address next.

\bigskip

\subsection{Exact expressions}

Four equations are needed to find the three quantities $\Lambda
_{m}=q_{m}(\sigma ^{-})/\sigma ^{2}$ $(m=0,1,2)$, as well as the parameter $%
K\left( \eta ,t,\alpha \right) $. We stress that the \textit{almost fully
analytical} determination of these unknown parameters was lacking in Ref.
[32] and represents an important part of the present work. Our detailed
analysis is given in Appendix B, and we quote here the main results.

i) For $\Lambda _{0}$, the same PY equation found by Baxter for isotropic
SHS \cite{Baxter68,Baxter71} 
\begin{equation}
12\eta t\ \Lambda _{0}^{2}-\left( 1+\frac{12\eta }{1-\eta }t\right) \Lambda
_{0}+y_{\sigma }^{\mathrm{HS}}(\eta )t=0.  \label{b5}
\end{equation}%
Only the smaller of the two real solutions (when they exist) is physically
significant \cite{Baxter68,Baxter71}, and reads

\begin{equation}
\Lambda _{0}=y_{0}^{\mathrm{PY}}(\sigma )t=\frac{y_{\sigma }^{\mathrm{HS}%
}(\eta )t}{\frac{1}{2}\left[ 1+\frac{12\eta }{1-\eta }t+\sqrt{\left( 1+\frac{%
12\eta }{1-\eta }t\right) ^{2}-48\eta \ y_{\sigma }^{\mathrm{HS}}(\eta )\
t^{2}}\right] },  \label{p1}
\end{equation}

ii) For $\Lambda _{1}$ and $\Lambda _{2}$, two other quadratic equations,
i.e. \ \ \ \ 
\begin{equation}
12\eta _{m}t\ \Lambda _{m}^{2}-\left( 1+\frac{12\eta _{m}}{1-\eta _{m}}%
t\right) \Lambda _{m}+h_{\sigma }^{\mathrm{HS}}(\eta _{m})t=-\mathcal{P}%
\text{\quad ~~}(m=1,2).  \label{p1b}
\end{equation}

iii) The fourth equation is the following linear relationship between $%
\Lambda _{1}$ and $\Lambda _{2}$%
\begin{equation}
\frac{12\eta _{2}\ \Lambda _{2}\ }{1-\eta _{2}}-\frac{12\eta _{1}\ \Lambda
_{1}\ }{1-\eta _{1}}=\frac{\eta _{2}\left( 4-\eta _{2}\right) \ }{\left(
1-\eta _{2}\right) ^{2}}-\frac{\eta _{1}\left( 4-\eta _{1}\right) \ }{\left(
1-\eta _{1}\right) ^{2}},  \label{p1c}
\end{equation}%
$\ $which stems from the condition $a_{2}=a_{1}$.

The analysis of Appendix B gives

\begin{equation}
\Lambda _{2}\left( \eta _{1},\eta _{2},t,\alpha \right) =\Lambda _{1}\left(
\eta _{2},\eta _{1},t,\alpha \right)
\end{equation}%
with

\begin{equation}
\Lambda _{m}=\Lambda +\Lambda _{m}^{\mathrm{ex}}\text{ \ \ \ \ \ \ \ }(m=1,2)
\label{p2a}
\end{equation}%
\begin{equation}
\Lambda =\frac{1}{3}+\frac{1}{4}\left( \frac{\eta _{1}}{1-\eta _{1}}+\frac{%
\eta _{2}}{1-\eta _{2}}\right) =\frac{1}{3}+\allowbreak \frac{x(1+4x)}{%
4\left( 1+x\right) \left( 1-2x\right) }  \label{p2b}
\end{equation}%
\begin{equation}
\Lambda _{1}^{\mathrm{ex}}=\frac{\eta _{2}}{4\left( 1-\eta _{2}\right) }%
W_{0}^{\mathrm{ex}},\qquad \Lambda _{2}^{\mathrm{ex}}=\frac{\eta _{1}}{%
4\left( 1-\eta _{1}\right) }W_{0}^{\mathrm{ex}},  \label{p2c}
\end{equation}%
where we have introduced $\eta _{1}=-x$, $\eta _{2}=2x$ \ ( $x\equiv K\eta $
), and $W_{0}^{\mathrm{ex}}$ is defined in Appendix B. All these quantites
are analytic functions of $x=K\eta $. Thus, to complete the solution, we
need an equation for $K$, which can be written as

\begin{equation}
K=\alpha t\ \mathcal{K},\text{ }\ \ \ \ \ \text{with \ \ \ }\ \mathcal{K}%
\text{ }=\frac{y_{0}^{\mathrm{PY}}(\sigma )}{Z(\eta _{1},\eta _{2},t)},
\label{p3}
\end{equation}%
\begin{equation}
Z=\frac{3}{2}\left( \Lambda _{1}+\Lambda _{2}\right) -3\left\{ \frac{1}{2}%
\sum_{m=1}^{2}\ \left[ 12\eta _{m}\ \Lambda _{m}^{2}-\frac{12\eta
_{m}\Lambda _{m}}{1-\eta _{m}}+h_{\sigma }^{\mathrm{HS}}(\eta _{m})\right] +%
\frac{K_{\mathrm{reg}}}{K}\right\} t  \label{p4}
\end{equation}%
and $\lim_{\eta \rightarrow 0}Z(\eta _{1},\eta _{2},t)=1$. Insertion of
found expressions for $\Lambda _{1},$ $\Lambda _{2}$ and $K_{\mathrm{reg}}$
(see Appendix B) into Eq. (\ref{p3}) yields a single equation for $K$ that
we have solved numerically, although some further analytic simplifications
are probably possible.

Our solution is then almost fully analytical, as only the final equation for 
$K$ is left to be solved numerically.

\subsection{Approximate expressions}

For practical use we next derive very accurate analytical approximations to $%
K$, $\Lambda _{1}$ and $\Lambda _{2}$, which provide an useful tool for
fully analytical calculations. Since in all cases of our interest we always
find $x=K\eta \ll 1$, a serie expansion leads to:

\begin{equation}
W_{0}^{\mathrm{ex}}=\frac{2}{3}\allowbreak \left( 1+5x\right) t+\mathcal{O}%
\left( x^{2}\right) ,
\end{equation}%
and, consequently,

\begin{equation}
\Lambda _{1}^{\mathrm{ex}}=\frac{x\left( 1+5x\right) }{3\left( 1-2x\right) }%
t+\mathcal{O}\left( x^{3}\right) ,\qquad \Lambda _{2}^{\mathrm{ex}}=-\frac{%
x\left( 1+5x\right) }{6\left( 1+x\right) }t+\mathcal{O}\left( x^{3}\right) .
\label{r5}
\end{equation}%
Similarly we can expand $Z$ in Eq. (\ref{p4}) as

\begin{equation}
Z(x,t)=1+z_{1}(t)x+z_{2}(t)x^{2}+O\left( x^{3}\right) ,
\end{equation}%
with%
\begin{equation}
\left\{ 
\begin{array}{c}
z_{1}(t)=\frac{1}{4}\left( 3+11t\right) \text{ \ \ \ \ \ \ \ \ \ } \\ 
z_{2}(t)=\frac{1}{4}\left( 15+61t-4t^{2}\right) .%
\end{array}%
\right.  \label{r6}
\end{equation}%
Insertion of this result into Eq. (\ref{p3}) yields a cubic equation for $K,$%
\begin{equation*}
z_{2}(t)\eta ^{2}K^{3}+z_{1}(t)\eta K^{2}+K-\alpha t\ y_{0}^{\mathrm{PY}%
}(\sigma )=0,
\end{equation*}%
which, again with the help of Eq. (\ref{p3}), is equivalent to a cubic
equation for $Z$ 
\begin{equation}
Z^{3}-Z^{2}+z_{1}(t)\left[ \alpha t\ y_{0}^{\mathrm{PY}}(\sigma )\eta \right]
Z+z_{2}(t)\left[ \alpha t\ y_{0}^{\mathrm{PY}}(\sigma )\eta \right] ^{2}=0.
\label{r7}
\end{equation}%
The physically acceptable solution then reads%
\begin{equation}
Z(\eta ,t)=\frac{1}{3}\left( 1+\sqrt[3]{\mathcal{B}+\sqrt{\mathcal{B}^{2}-%
\mathcal{C}^{3}}}+\sqrt[3]{\mathcal{B}-\sqrt{\mathcal{B}^{2}-\mathcal{C}^{3}}%
}\right) ,  \label{r8}
\end{equation}%
where

\begin{equation}
\left\{ 
\begin{array}{c}
\mathcal{B}=1+\frac{9}{2}z_{1}(t)\left[ \alpha t\ y_{0}^{\mathrm{PY}}(\sigma
)\eta \right] +\frac{27}{2}z_{2}(t)\left[ \alpha t\ y_{0}^{\mathrm{PY}%
}(\sigma )\eta \right] ^{2} \\ 
\mathcal{C}=1+3z_{1}(t)\left[ \alpha t\ y_{0}^{\mathrm{PY}}(\sigma )\eta %
\right] .\text{ \ \ \ \ \ \ \ \ \ \ \ \ \ \ \ \ \ \ \ \ \ \ \ \ \ \ \ \ \ }%
\end{array}%
\right.  \label{r9}
\end{equation}%
In conclusion, our approximate analytic solution for $K$, $\Lambda _{1}$ and 
$\Lambda _{2}$ includes three simple steps: i) calculate $K$ by using Eqs. (%
\ref{p3}), (\ref{r8})-(\ref{r9}), (\ref{r6}); ii) evaluate $x=K\eta $; iii)
solve for $\Lambda _{1}$ and $\Lambda _{2}$ by means of Eqs. (\ref{p2b}) and
(\ref{r5}).

\bigskip

\subsection{Numerical comparison}

In order to assess the precision of previous approximations, we have
calculated $K$, $\Lambda _{1}$ and $\Lambda _{2}$ by two methods: i) solving
numerically Eqs. (\ref{s2}), and ii) using our analytic approximations.
After fixing $\alpha =1/2,$ we have increased the adhesion strength (or
decreased the temperature) from $t=0$ (HS limit) up to $t=0.8$, for some
representative values of the volume fraction ($\eta =0.01$, $0.1$, $0.2$ and 
$0.4)$. The maximum value of $t$ corresponds to $\tau =1/(12t)\simeq 0.1$,
which lies close to the critical temperature of the isotropic SHS fluid. On
the other hand, $\eta =0.01$ has been chosen to illustrate the fact that, as 
$\eta \rightarrow 0$, the parameter $K$ tends to $\alpha t$. The linear
dependence of $K$ on $t$ in this case is clearly visible in the top panel of
Figure 2.

In Figures 2 and 3 the exact and approximate results for $K$, $\Lambda _{1}$
and $\Lambda _{2}$ are compared. The agreement is excellent: at $\eta =0.1$, 
$0.2$ and $0.4$, the relative error on $K$ does not exceed $0.1\%$, $0.4\%$
and $1\%$, respectively, while the maximum of the absolute relative errors
on $\Lambda _{1}$ and $\Lambda _{2}$ always remain less than $0.05$, $0.2$
and $0.6~\%$ in the three above-mentioned cases. It is worth noting that, as 
$\eta $ increases, the variations of $\Lambda _{1}$ and $\Lambda _{2}$ are
always relatively small; on the contrary, $K$ experiences a marked change,
with a progressive lowering of the relevant curve.

\bigskip

\section{Some illustrative results on the local orientational structure}

Armed with the knowledge of the analytic expression for the $q_{m}s$ a rapid
numerical calculation of the three harmonic coefficients $\left\{
h_{0},h_{\Delta },h_{D}\right\} $ appearing in

\begin{equation}
g^{\mathrm{PY-OL}}(1,2)=1+h_{0}(r)+h_{\Delta }(r)\Delta (1,2)+h_{D}(r)D(1,2).
\label{g1}
\end{equation}%
can be easily obtained as follows. From the second Baxter IE ($\ref{ie2b}$),
one can generate $h(r)$ directly from $q(r)$, avoiding the passage through $%
c(r)$. From $\left\{ q_{0},q_{1},q_{2}\right\} $ one first obtains $\left\{
h_{0},h_{1},h_{2}\right\} $, by applying a slight extension of Perram's
numerical method \cite{Perram75} and then derive $\left\{ h_{0},h_{\Delta
},h_{D}\right\} $, according to the above-mentioned recipes.

The main aim of the present paper was to present the necessary mathematical
machinery to investigate thermophysical properties. We now illustrate the
interest of the model by reporting some preliminary numerical results on the
orientational dependence of $g^{\mathrm{PY-OL}}(1,2) $ -- i.e. on the local
orientational structure -- as a consequence of the anisotropic adhesion. A
more detailed analysis will be reported in a forthcoming paper.

Consider the configuration depicted in Figure 4. Let a generic particle $1$
be fixed at a position $\mathbf{r}_{1\text{ }}$in the fluid with orientation 
$\mathbf{u}_{1\text{ }}$, and consider another particle $2$ located along
the straigth half-line which originates from the center of $1$ and with
direction $\mathbf{u}_{1\text{ }}$. This second particle has then a fixed
distance $r$ from $1$, but can assume all possible orientations $\mathbf{u}%
_{2\text{ }}$, which -- by axial symmetry -- can be described by a single
polar angle $\theta \equiv \theta _{2}$ (i.e., the angle between $\mathbf{u}%
_{1\text{ }}$and $\mathbf{u}_{2\text{ }}$) with respect to the
intermolecular reference frame. Within this geometry, we have $\left( \theta
_{1},\varphi _{1}\right) =(0,0)$ and $\varphi _{2}=0$, obtaining $\Delta
(1,2)=\cos \theta $, $D(1,2)=2\cos \theta $. Consequently, $%
g(1,2)=g(r,\theta _{1},\varphi _{1},\theta _{2},\varphi _{2})$ reduces to 
\begin{equation}
g(r,\theta )=g_{0}(r)+\left[ h_{\Delta }(r)+2h_{D}(r)\right] \cos \theta ,
\label{eq6}
\end{equation}%
where $\theta \equiv \theta _{2}$, and $g_{0}(r)=1+h_{0}(r)$ is the radial
distribution function of the reference isotropic SHS fluid.

Clearly, $g(r,\theta )$ is proportional to the probability of finding, at a
distance $r$ from a given molecule $1$, a molecule $2$ having a \textit{%
relative} orientation $\theta $. We consider the three most significant
values of this angle: i) $\theta =0$, which corresponds to the `parallel'
configuration of $\mathbf{u}_{1}$ and $\mathbf{u}_{2}$; ii) $\theta =\pi /2$%
, for the `orthogonal' configuration; and $\theta =\pi $, for the two
`antiparallel' (head-to-head and tail-to-tail) configurations. From Eq. (\ref%
{eq6}) it follows that%
\begin{equation}
\begin{array}{c}
g^{\mathrm{par}}(r)=g(r,0)=g_{0}(r)+\left[ h_{\Delta }(r)+2h_{D}(r)\right] ,%
\text{ \ \ \ } \\ 
g^{\mathrm{ortho}}(r)=g(r,\pi /2)=g_{0}(r),\text{ \ \ \ \ \ \ \ \ \ \ \ \ \
\ \ \ \ \ \ \ \ \ \ \ } \\ 
g^{\mathrm{antipar}}(r)=g(r,\pi )=g_{0}(r)-\left[ h_{\Delta }(r)+2h_{D}(r)%
\right] .\text{ \ }%
\end{array}
\label{g2}
\end{equation}%
Note that $g^{\mathrm{ortho}}(r)$ coincides with the isotropic result $%
g_{0}(r)$.

In Figures 5 we depict the above sections through the three-dimensional
surface corresponding to $g(r,\theta )$, i.e., $g^{\mathrm{par}}(r)$, $g^{%
\mathrm{ortho}}(r)$ and $g^{\mathrm{antipar}}(r)$, for $\eta =0.3$ with $%
t=0.2$ and $t=0.6$, respectively, at the highest asymmetry value admissible
in the present model, i.e. $\alpha =1/2$. The most significant features from
these plots are: i) $g^{\mathrm{antipar}}(\sigma ^{+})>g^{\mathrm{par}%
}(\sigma ^{+})$; ii) for $r>2\sigma $ $g^{\mathrm{antipar}}(r)\approx g^{%
\mathrm{par}}(r)\approx g_{0}(r)$, i.e., the anisotropic adhesion seems to
affect only the first coordination layer, $\sigma <r<2\sigma $, around each
particle.

The interpretation of these results is the following. In view of i) we see
that the parallel configuration is less probable than the antiparallel one
at contact. Such a finding, together with ii), means that chain formation
characteristic of polymerization is inhibited by the short-ranged
anisotropic adhesion exploited here. This strictly contrasts with the case
of long-ranged DHS fluids, where it is believed \cite{Camp00,Tlusty00} that
chaining phenomena might preempt the gas-liquid transition. This specific
feature of the present model is extremely interesting and we plan a
throughout investigation on this topic in a future publication.

\bigskip

\section{Phase stability}

In view of previous findings, a very natural question is whether the
addition of our anisotropic sticky term to the potential changes phase
stability and phase transition curves with respect to the corresponding
isotropic case. We believe the answer to be positive. This is strongly
suggested by results obtained for similar anisotropic models, such as hard
spheres with `sticky points'\ \cite%
{Sciortino05,Bianchi06,Michele06,Lomakin99,Starr03,Zhang04,Glotzer04a,Glotzer04b,Sciortino07}
or `sticky patches'\ \cite%
{Jackson88,Ghonasgi95,Sear99,Mileva00,Kern03,Zhang05,Fantoni07}.

We now briefly comment on this issue. Within our formalism, this problem of
stability can be conveniently analyzed using standard formalism
devised to this aim \cite{Gray84,Stecki81,Chen92,Klapp97}.

We start from the stability condition with respect to small but arbitrary
fluctuations of the one-particle density $\rho(1)$ from the equilibrium configuration,
denoted as 'eq' \cite{Stecki81,Chen92,Klapp97}, 

\begin{eqnarray}
\int d(1) \int d(2) \left [\frac{\delta(1,2)}{\rho(1)} - c\left(1,2\right) 
\right]_{eq} \delta \rho(1) \delta \rho(2) > 0.
\label{stability}
\end{eqnarray}

Here $d(i)$ stands for $d \mathbf{r}_i ~d \Omega_i$, $i=1,2$, and we assume the
equilibrium one-particle density to be $\rho/4\pi$ \cite{Stecki81,Chen92,Klapp97}. 

We expand the fluctuations both in Fourier modes and in spherical harmonics
\cite{Gray84}

\begin{eqnarray}
\delta \rho \left(j\right) \equiv \delta \rho \left(\mathbf{r}_j,\Omega_j\right)
&=& \int \frac{d \mathbf{k}}{\left(2\pi\right)^3} ~\mathrm{e}^{\mathrm{i} \mathbf{k} \cdot
\mathbf{r}_j} \sum_{l=0}^{+\infty} \sum_{m=-l}^{+l} \delta \tilde{\rho} \left(\mathbf{k}
\right) Y_{lm} \left(\Omega_j\right). 
\label{expansion}
\end{eqnarray}

Using the orthogonality relation \cite{Gray84}

\begin{eqnarray}
\int d \Omega ~Y_{lm}^{*} \left(\Omega\right) Y_{l'm'}\left(\Omega\right) &=&
\delta_{l l'} \delta_{m m'},
\label{orthogonality}
\end{eqnarray}

standard manipulations \cite{Klapp97} show that condition (\ref{stability}) can
be recast into the form

\begin{eqnarray}
\sum_{l_{1},l_{2}=0}^{+\infty} \sum_{m_{1}=-l_{1}}^{+l_{1}} \sum_{m_{2}=-l_{2}}^{+l_{2}}
\int \frac{d \mathbf{k}}{\left(2 \pi\right)^3} ~ \delta \tilde{\rho}_{l_{1} m_{1}} 
\left(\mathbf{k} \right) \delta \tilde{\rho}_{l_{1} m_{1}}^{*} 
\left(\mathbf{k} \right) \tilde{A}_{l_{1} m_{1} l_{2} m_{2}} \left(\mathbf{k} \right) &>0&,
\label{stability2}
\end{eqnarray}

where the matrix elements $\tilde{A}_{l_{1} m_{1} l_{2} m_{2}} \left(\mathbf{k} \right)$
are given by

\begin{eqnarray} \label{matrix}
\tilde{A}_{l_{1} m_{1} l_{2} m_{2}} \left(\mathbf{k} \right) &=& \left(-1\right)^{m_{1}}
\frac{4 \pi}{\rho} \delta_{l_{1} l_{2}} \delta_{m_{1},-m_{2}} -
\int d \Omega_1 \int d ~ \Omega_{2} Y_{l_{1} m_{1}} \left(\Omega_1\right) 
Y_{l_{2} m_{2}} \left(\Omega_2\right) \\ \nonumber
&\times&  \int d\mathbf{r} ~\mathrm{e}^{\mathrm{i} \mathbf{k}
\cdot \mathbf{r}} c\left(\mathbf{r},\Omega_1,\Omega_2\right).
\end{eqnarray}

The problem of the stability has been reported to the character 
of the eigenvalues of matrix (\ref{matrix}). This turns out to be particularly
simple in our case. Using the results (\ref{wertheim_int}) it is easy to
see that

\begin{eqnarray}
\int d \mathbf{r}~ \mathrm{e}^{\mathrm{i} \mathbf{k} \cdot \mathbf{r}}
c\left(\mathbf{r},\Omega_1,\Omega_2 \right) &=& \tilde{c}_{0} \left(k\right)
+\tilde{c}_{\Delta} \left(k\right) \Delta\left(\Omega_1,\Omega_2\right) +
\overline{c}_{D} \left(k\right) D\left(\Omega_1,\Omega_2,\Omega_k\right).
\label{integral_c}
\end{eqnarray}

Insertion of Eq.(\ref{integral_c}) into Eq.(\ref{matrix}) leads to

\begin{eqnarray} \label{matrix2}
\tilde{A}_{l_{1} m_{1} l_{2} m_{2}} \left(\mathbf{k} \right) &=& 
\left(-1\right)^{m_{1}}
\frac{4 \pi}{\rho} \delta_{l_{1} l_{2}} \delta_{m_{1},-m_{2}} \\ \nonumber
&-& \left[
\tilde{c}_{0}\left(k\right) I_{l_{1} m_{1} l_{2} m_{2}}^{(0)}
+\tilde{c}_{\Delta} \left(k\right) I_{l_{1} m_{1} l_{2} m_{2}}^{(\Delta)}
+\tilde{c}_{D} \left(k\right) I_{l_{1} m_{1} l_{2} m_{2}}^{(D)}
\right],
\end{eqnarray}

\noindent where we have introduced the following integrals, which can be evaluated
in the intermolecular frame, using standard properties of the
spherical harmonics \cite{Gray84}

\begin{eqnarray} \label{integrals}
I_{l_{1} m_{1} l_{2} m_{2}}^{(0)} &\equiv& \int d \Omega_1 \int d \Omega_2
~Y_{l_{1} m_{1}} \left(\Omega_1\right) Y_{l_{2} m_{2}} \left(\Omega_2
\right) = 4 \pi \delta_{l_{1} 0,} \delta_{l_{2},0} \delta_{m_{1} 0}
\delta_{m_{2}} \delta_{0} \\ \nonumber
I_{l_{1} m_{1} l_{2} m_{2}}^{(\Delta)} &\equiv& \int d \Omega_1 \int d \Omega_2
~Y_{l_{1} m_{1}} \left(\Omega_1\right) Y_{l_{2} m_{2}} \left(\Omega_2
\right) ~\Delta\left(\Omega_1,\Omega_2\right)= \frac{4}{3} \pi \delta_{l_{1} 1} 
\delta_{l_{2},1} \delta_{m_{1} 0} \delta_{m_{2},0} \\ \nonumber
I_{l_{1} m_{1} l_{2} m_{2}}^{(D)} \left(\cos \theta \right) &\equiv& 
\int d \Omega_1 \int d \Omega_2
~Y_{l_{1} m_{1}} \left(\Omega_1\right) Y_{l_{2} m_{2}} \left(\Omega_2
\right) D\left(\Omega_1,\Omega_2,\Omega_k\right) \\ \nonumber 
&=& \frac{4}{3} \pi \delta_{l_{1},1} 
\delta_{l_{2},1} \delta_{m_{1}, 0}
\delta_{m_{2},0} ~2 ~P_{2}\left(\cos \theta \right) 
\end{eqnarray}

and where $P_2(x)=(3 x^2 -1)/2$ is the second Legendre polynomial.

Hence, the matrix (\ref{matrix}) is diagonal and the relevant terms
are

\begin{eqnarray}
\tilde{A}_{0000} \left(k\right) &=& 4 \pi
\left[ \frac{1}{\rho}-\tilde{c}_{0} \left(k\right) \right],
\label{element00}
\end{eqnarray}
whose positiveness is recognized as the isotropic stability condition,
and
\begin{eqnarray}
\tilde{A}_{1010}\left(\mathbf{k}\right) &=& 4 \pi \left\{
\frac{1}{\rho} -\frac{1}{3} \left[\tilde{c}_{\Delta}\left(k\right)
+ 2 P_{2} \left(\cos \theta\right) \overline{c}_{D}\left(k\right)
\right] \right\}.
\label{element11}
\end{eqnarray} 

\noindent All remaining diagonal terms have the form $\tilde{A}_{l0l0}=
4 \pi/\rho>0$. 

In order to test for possible angular instabilities, we consider
the limit $k \to 0$ of Eq.~(\ref{element11}) namely

\begin{eqnarray}
\tilde{A}_{1010}\left(0\right) &=& \frac{4 \pi}{\rho} \left\{
1 -\frac{\rho}{3} \left[\tilde{c}_{\Delta}\left(0\right)
+ 2 P_{2} \left(\cos \theta\right) \overline{c}_{D}\left(0\right)
\right] \right\}.
\label{element11k0}
\end{eqnarray} 
\noindent
This can be quickly computed with the aid of Eqs.~(\ref{so11}), (\ref{bf1}),
the fact that $\bar{c}_{D}(0)=\tilde{c}_{D}^{0}(0)$ and the identity
(\ref{so10}). We find

\begin{eqnarray}
\label{element11_res}
\tilde{A}_{1010}\left(0\right) &=& \frac{4 \pi}{\rho} a_1^2,
\end{eqnarray}
which is independent of the angle $\theta$. This value is found to be always
positive as $a_1>0$ (see Fig.\ref{fig6}).
Within this first-order approximation, therefore the only instability
in the system stems from the isotropic compressibility. 
The reason for this can be clearly traced back to the first-order
approximation to the angular dependence of the correlation functions. If
quadratic terms in $\Delta $ and $D$ were included into the series expansion
for correlation functions, the particular combination leading to a cancellation
of the angular dependence in the stability matrix 
$\tilde{A}_{l_{1}m_{1}l_{2}m_{2}}\left(0\right)$ would not occur, leading to
a different result. 

This fact is consistent with the more general statement that, in any
approximate theory, thermodynamics usually requires a higher degree of
theoretical accuracy than the one sufficient for obtaining significant
structural data. Conceptually, the need of distinguishing structural results
from thermodynamical ones is rather common. For instance, in statistical
mechanics of liquids it is known that approximating the model potential only
with its repulsive part (for instance, the hard sphere term) can account for
all essential features of the structure, but yields unsatisfactory
thermodynamics. On the other hand, the present paper refers to a \textit{%
simplified} statistical-mechanical tool, i.e. the OZ equation within our
PY-OL closure, which has been explicitly selected to allow an analytical
solution. Our results however indicate that the first-order expansion used
in the PY-OL closure can give reasonable information about structure, but
not on thermodynamics, where a higher level of sophistication is required.

\section{Concluding remarks}

In this paper we have discussed an anisotropic variation of the original
Baxter model of hard spheres with surface adhesion. In addition to the HS
potential, molecules of the fluid interact via an isotropic sticky
attraction plus an additional anisotropic sticky correction, whose strength
depends on the orientations of the particles in dipolar way. By varying the
value of a parameter $\alpha $, the anisotropy degree can be changed.
Consequenly, the strength of the total sticky potential can vary from twice
the isotropic one down to the limit of no adhesion (HS limit). These
particles may be regarded as having two non-uniform, hemispherical,
`dipolar-like patches', thus providing a link with uniformly adhesive
patches \cite{Jackson88,Ghonasgi95,Sear99,Mileva00,Kern03,Zhang05,Fantoni07}.

We have obtained a full analytic solution of the molecular OZ equation,
within the PY-OL approximation, by using Wertheim's technique \cite%
{Wertheim71}. Our PY-OL approximation should be tested against exact
computer simulations, in order to assess its reliability. Nevertheless, we
may reasonably expect the results to be reliable even at experimentally
significant densities, notwithstanding the truncation of the higher-orders
terms in the angular expansion. Only one equation, for the parameter $K$,
has to be solved numerically. In additon, we have provided analytic
approximations to $K$, $\Lambda _{1}$ and $\Lambda _{2}$ so accurate that,
in practice, the whole solution can really be regarded as fully analytical.
From this point of view, the present paper complements the above-mentioned
previous work by Blum \textit{et al. }\cite{Blum90}.

We have also seen that thermophysical properties require a more detailed
treatment of the angular part than the PY-OL closure. Nonetheless, 
even within the PY-OL oversimplified framework, our findings
are suggestive of a dependence of the fluid-fluid coexistence line on
anisotropy.

Our analysis envisions a number of interesting perspectives, already hinted
by the preliminary numerical results reported here. It would be very
interesting to compare the structural and thermodynamical properties of this
model with those stemming from truly dipolar hard spheres \cite{Stecki81,
Chen92,Klapp97}.
The possibility of local orientational ordering can be assessed by computing
the pair correlation function $g(1,2)$ for the most significant
interparticle orientations. We have shown that this task can be easily
performed within our scheme. This should provide important information about
possible chain formation and its subtle interplay with the location of the
fluid-fluid transition line. The latter bears a particular interest in view
of the fact that computer simulations on DHS are notoriously difficult and
their predictions regarding the location of such a transition line have
proven so far unconclusive \cite{Frenkel02}. The long-range nature of DHS
interactions may in fact promote polymerization preempting the usual
liquid-gas transition \cite{Tlusty00}. Our preliminary results on the
present model strongly suggest that this is not the case for sufficiently
short-ranged interactions, thus allowing the location of such a transition
line to be studied as a function of the anisotropy degree of the model. Our
sticky interactions have only attractive adhesion, the only repulsive part
being that pertinent to hard spheres, whereas the DHS potential is both
attractive and repulsive, depending on the orientations.

Finally, information about the structural ordering in the present model
would neatly complement those obtained by us in a recent parallel study on a
SHS fluid with one or two uniform circular patches \cite{Fantoni07}. Work
along this line is in progress and will be reported elsewhere.

\acknowledgments 
We acknowledge financial support from PRIN 2005027330. It is our pleasure to
thank Giorgio Pastore and Mark Miller for enlighting discussions on the
subject.

\bigskip \appendix

\section{Extension of Wertheim's approach}

The Fourier transform of the \textit{excess PY-OL equation}, Eq. (\ref{ozeq2}%
), reads

\begin{equation}
\widehat{h}_{\mathrm{ex}}(\mathbf{k},\Omega _{1},\Omega _{2})=\widehat{c}_{%
\mathrm{ex}}(\mathbf{k},\Omega _{1},\Omega _{2})+\rho \ \left\langle \ 
\widehat{c}_{\mathrm{ex}}(\mathbf{k},\Omega _{1},\Omega _{3})\ \widehat{h}_{%
\mathrm{ex}}(\mathbf{k},\Omega _{3},\Omega _{2})\ \right\rangle _{\Omega
_{3}}  \label{w3}
\end{equation}%
(the superscripts have been omitted for simplicity). In order to evaluate
the angular average, we first need the FT of $c$ and $h$. The FT integral (%
\ref{oz5}) may be rewritten as

\begin{equation*}
\int_{0}^{\infty }dr\ r^{2}\int d\Omega _{r}\ \exp (i\mathbf{k\cdot r})\ %
\left[ \cdots \right] =\int_{0}^{\infty }dr\ r^{2}\int_{0}^{2\pi }d\phi
\int_{-1}^{+1}d\left( \cos \theta \right) \ e^{ikr\cos \theta }\ \left[
\cdots \right] .
\end{equation*}%
Let us now apply this operator to $F_{\mathrm{ex}}(1,2)$ ($F=c,h$),
expressed as

\begin{equation}
F_{\mathrm{ex}}(\mathbf{r},\Omega _{1},\Omega _{2})=F_{\Delta }(r)\ \Delta
(\Omega _{1},\Omega _{2})+F_{D}(r)\ D(\Omega _{1},\Omega _{2},\Omega _{r}),
\end{equation}%
and first perform the angular integration $\int d\Omega _{r}$, recalling
that \cite{Wertheim71}

\begin{equation} \label{wertheim_int}
\begin{array}{c}
\int d\Omega _{r} ~ \exp (i\mathbf{k\cdot r})\ \ 1=4\pi \ j_{0}(kr)\ 1\text{
\ \ \ } \\ 
\int d\Omega _{r}\ \exp (i\mathbf{k\cdot r})\ \ \Delta (\Omega _{1},\Omega
_{2})=4\pi \ j_{0}(kr)\ \Delta (\Omega _{1},\Omega _{2})\text{\ \ \ \ } \\ 
\int d\Omega _{r}\ \exp (i\mathbf{k\cdot r})\ \ D(\Omega _{1},\Omega
_{2},\Omega _{r})=-4\pi \ j_{2}(kr)\ D(\Omega _{1},\Omega _{2},\Omega _{k}),
\\ 
\end{array}%
\end{equation}%
where $j_{0}(x)=x^{-1}\sin x$ and $j_{2}(x)=3x^{-3}\sin x-3x^{-2}\cos
x-j_{0}(x)$ are Bessel functions, and 
\begin{equation*}
D(\Omega _{1},\Omega _{2},\Omega _{k})=3(\mathbf{u}_{1}\cdot \hat{\mathbf{k}}%
)(\mathbf{u}_{2}\cdot \hat{\mathbf{k}})-\mathbf{u}_{1}\cdot \mathbf{u}%
_{2}\equiv D_{k}(1,2),
\end{equation*}%
with $\hat{\mathbf{k}}=\mathbf{k}/k$. We get

\begin{equation*}
\widehat{F}_{\mathrm{ex}}(\mathbf{k},\Omega _{1},\Omega _{2})=\widetilde{F}%
_{\Delta }(k)\ \Delta (\Omega _{1},\Omega _{2})+\overline{F}_{D}(k)\
D(\Omega _{1},\Omega _{2},\Omega _{k}),
\end{equation*}%
where $\widetilde{F}_{\Delta }(k)$ is the usual FT of the spherically
symmetric function $F_{\Delta }(r)$: $\ \widetilde{F}_{\ldots }(k)=4\pi
\int_{0}^{\infty }dx\ x^{2}\ j_{0}(kx)F_{\ldots }(x)$. On the other hand, $%
\overline{F}_{D}(k)=-4\pi \int_{0}^{\infty }dx\ x^{2}\ j_{2}(kx)F_{D}(x)$,
which is the Hankel transform of $F_{D}(r)$, may conveniently be considered
as the FT of a `modified' function $F_{D}^{0}(r)$, i.e. $\overline{F}_{D}(k)=%
\widetilde{F}_{D}^{0}(k)$. Taking the inverse FT of $\overline{F}_{D}(k)$
yields 
\begin{equation}
F_{D}^{0}(r)=\frac{1}{2\pi ^{2}}\int_{0}^{\infty }dk\ k^{2}\ j_{0}(kr)%
\overline{F}_{D}(k)=F_{D}(r)-3\int_{r}^{\infty }\frac{F_{D}(x)}{x}\ dx,
\end{equation}%
with the help of the identity$\ $%
\begin{equation*}
\int_{0}^{\infty }dk\ k^{2}\ j_{0}(kr)\ j_{2}(kx)=\frac{\pi }{2}\left[ \frac{%
3\theta \left( x-r\right) }{x^{3}}-\frac{\delta \left( x-r\right) }{x^{2}}%
\right] .
\end{equation*}

In conclusion, the FT of $F_{\mathrm{ex}}(1,2)$ reads 
\begin{equation}
\widehat{F}_{\mathrm{ex}}(\mathbf{k},\Omega _{1},\Omega _{2})=\widetilde{F}%
_{\Delta }(k)\ \Delta (\Omega _{1},\Omega _{2})+\widetilde{F}_{D}^{0}(k)\
D(\Omega _{1},\Omega _{2},\Omega _{k}),  \label{oz12c}
\end{equation}%
with $F$ standing for $h$ or $c$.

Let us now define the \textit{angular convolution} of two functions as

\begin{equation*}
A\circ B=B\circ A\equiv \left\langle A(\Omega _{1},\Omega _{3})\ B(\Omega
_{3},\Omega _{2})\right\rangle _{\Omega _{3}}.\ 
\end{equation*}%
Wertheim \cite{Wertheim71} demonstrated that the rotational invariants $1,$ $%
\Delta ,$ and $D$ form a closed group under angular convolution; that is,
the angular convolution of any two members of this set yields only a
function in the same set, or zero, according to the following multiplication
table

\bigskip

\begin{table}[h]
\begin{tabular}{lcccccc}
\hline
$\circ$ & $~~~~~~~$ & $1$ & $~~~~~~~$ & $\Delta$ & $~~~~~~~$ & $D_{k}$ \\ 
\hline
$1$ &  & $1$ &  & $0$ &  & $0$ \\ 
$\Delta$ &  & $0$ &  & $\Delta/3$ &  & $D_{k}/3$ \\ 
$D_{k}$ &  & $0$ &  & $D_{k}/3$ &  & $(D_{k}+2\Delta)/3$ \\ \hline
\end{tabular}%
\caption{Angular convolutions of the basis functions $1,\Delta $ and $D_{k}$%
. }
\label{tab1:temp}
\end{table}
Substituting the expressions for $\widehat{c}_{\mathrm{ex}}$ and$\ \widehat{h%
}_{\mathrm{ex}}$ given by Eq. (\ref{oz12c}) into the angular average $%
\widehat{c}_{\mathrm{ex}}\circ \widehat{h}_{\mathrm{ex}}=\left\langle 
\widehat{c}_{\mathrm{ex}}(\mathbf{k},\Omega _{3},\Omega _{2})\ \widehat{h}_{%
\mathrm{ex}}(\mathbf{k},\Omega _{3},\Omega _{2})\right\rangle _{\Omega _{3}}$%
, with the help of Table I we obtain%
\begin{eqnarray*}
\widehat{c}_{\mathrm{ex}}\circ \widehat{h}_{\mathrm{ex}} &=&\widetilde{c}%
_{\Delta }\widetilde{h}_{\Delta }\frac{1}{3}\Delta +\widetilde{c}_{\Delta }%
\widetilde{h}_{D}^{0}\frac{1}{3}D_{k} \\
&&+\widetilde{c}_{D}^{0}\widetilde{h}_{\Delta }\frac{1}{3}D_{k}+\widetilde{c}%
_{D}^{0}\widetilde{h}_{D}^{0}\frac{1}{3}\left( 2\Delta +D_{k}\right) .
\end{eqnarray*}%
Inserting this result into Eq. (\ref{w3}) and equating the coefficients of $%
\Delta ,$ and $D$ separately, one finds that the $\mathbf{k}$-space excess
PY-OL equation splits into two coupled integral equations, i.e., 
\begin{equation*}
\left\{ 
\begin{array}{c}
\widetilde{h}_{\Delta }-\widetilde{c}_{\Delta }=\frac{1}{3}\rho ~\left( \ 
\widetilde{c}_{\Delta }\widetilde{h}_{\Delta }+2\ \widetilde{c}_{D}^{0}%
\widetilde{h}_{D}^{0}\ \right) \text{\ \ \ \ \ \ \ \ \ \ \ \ \ } \\ 
\widetilde{h}_{D}^{0}-\widetilde{c}_{D}^{0}=\frac{1}{3}\rho ~\left( \ 
\widetilde{c}_{\Delta }\widetilde{h}_{D}^{0}+\widetilde{c}_{D}^{0}\widetilde{%
h}_{\Delta }+\widetilde{c}_{D}^{0}\widetilde{h}_{D}^{0}\ \right) .\text{ \ \ 
}%
\end{array}%
\right.  \label{PYOL}
\end{equation*}%
Coming back to the $\mathbf{r}$-space, one gets the following equations

\begin{equation}
\left\{ 
\begin{array}{c}
h_{\Delta }(r)=c_{\Delta }(r)+\frac{1}{3}\rho ~\left( \ c_{\Delta }\star
h_{\Delta }+2\ c_{D}^{0}\star h_{D}^{0}\ \right) \text{\ \ \ \ \ \ \ \ \ \ \ 
} \\ 
h_{D}^{0}(r)=c_{D}^{0}(r)+\frac{1}{3}\rho ~\left( \ c_{\Delta }\star
h_{D}^{0}+c_{D}^{0}\star h_{\Delta }+c_{D}^{0}\star h_{D}^{0}\ \right) .%
\end{array}%
\right.  \label{oz11}
\end{equation}

In particular, since $h_{D}(r)=0$ for \ $0<r<\sigma $, Eq. (\ref{oz12})
yields $h_{D}^{0}(r)=-3K$ for \ $0<r<\sigma $, with $K$ being a
dimensionless parameter defined by 
\begin{equation}
K=\int_{\sigma ^{-}}^{\infty }\frac{h_{D}(x)}{x}\ dx.  \label{oz14}
\end{equation}

The exact core conditions for Eqs. (\ref{oz11}) are%
\begin{equation}
\left. 
\begin{array}{c}
\text{ \ }h_{\Delta }(r)=0\text{ \ \ \ \ \ \ } \\ 
h_{D}^{0}(r)=-3K%
\end{array}%
\right\} \text{ \ \ \ \ \ for \ }0<r<\sigma .  \label{oz12b}
\end{equation}%
Now, in the PY-OL closure for the DCFs, Eqs. (\ref{closure}), the closure
for $c_{D}(r)$ must be replaced with that corresponding to $c_{D}^{0}(r)$
(for simplicity, here and in the following we omit the superscript PY-OL).
In order to derive this, let us start from $c_{D}(r)=c_{D\text{,reg}%
}(r)+\Lambda _{D}\ \sigma \delta \left( r-\sigma \right) ,$ where $c_{D\text{%
,reg}}(r)=f^{\mathrm{HS}}(r)\ y_{D}^{\mathrm{PY}}(r)=0$ for $r\geq \sigma $.
Then Eq. (\ref{oz12}) yields%
\begin{equation*}
c_{D}^{0}(r)=c_{D}(r)-3\int_{r}^{\sigma }\frac{c_{D\text{,reg}}(x)}{x}\
dx-3\Lambda _{D}\ \theta (\sigma -r),
\end{equation*}%
since $\int_{r}^{\infty }\delta \left( x-\sigma \right) \ x^{-1}\ dx=\sigma
^{-1}\theta (\sigma -r)$ \cite{note1}. So we get%
\begin{equation}
c_{D}^{0}(r)=c_{D}(r)\text{ \ \ \ for \ }r\geq \sigma ,
\end{equation}%
and the required new closures are%
\begin{equation}
\left. 
\begin{array}{c}
c_{\Delta }(r)=\Lambda _{\Delta }\ \sigma \delta \left( r-\sigma \right) \\ 
c_{D}^{0}(r)=\Lambda _{D}\ \sigma \delta \left( r-\sigma \right)%
\end{array}%
\right\} \text{ \ \ \ \ \ }r\geq \sigma .  \label{c7}
\end{equation}

In order to \textit{decouple} the two integral equations for $\Delta $- and $%
D$-coefficients, we then introduce two new unknown functions, which are
linear combinations of the previous ones. Defining%
\begin{equation*}
\ \widetilde{F}_{\mathrm{new}}=\lambda _{1}\widetilde{F}_{\Delta }+\lambda
_{2}\widetilde{F}_{D}^{0}\text{ \ \ \ \ \ \ }\left( F=c,h\right)
\end{equation*}%
and using Eqs. (\ref{PYOL}) leads to%
\begin{eqnarray*}
\widetilde{h}_{\mathrm{new}}-\widetilde{c}_{\mathrm{new}} &=&\lambda
_{1}\left( \widetilde{h}_{\Delta }-\widetilde{c}_{\Delta }\right) +\lambda
_{2}\left( \widetilde{h}_{D}^{0}-\widetilde{c}_{D}^{0}\right) \\
&=&\frac{1}{3}\rho ~\left[ \ \lambda _{1}\widetilde{c}_{\Delta }\widetilde{h}%
_{\Delta }+\lambda _{2}\left( \widetilde{c}_{\Delta }\widetilde{h}_{D}^{0}+%
\widetilde{c}_{D}^{0}\widetilde{h}_{\Delta }\right) +\left( 2\lambda
_{1}+\lambda _{2}\right) \widetilde{c}_{D}^{0}\widetilde{h}_{D}^{0}\right] .
\end{eqnarray*}%
Requiring the second member of this equation to be proportional to $\rho 
\widetilde{c}_{\mathrm{new}}\widetilde{h}_{\mathrm{new}}$ -- that is, equal
to $\mathcal{L}\rho \left( \lambda _{1}\widetilde{c}_{\Delta }+\lambda _{2}%
\widetilde{c}_{D}^{0}\right) \left( \lambda _{1}\widetilde{h}_{\Delta
}+\lambda _{2}\widetilde{h}_{D}^{0}\right) $, with $\mathcal{L}$ being the
proportionality constant --, yields the following conditions%
\begin{equation*}
\left\{ 
\begin{array}{c}
\frac{1}{3}\lambda _{1}=\mathcal{L}\lambda _{1}^{2}\text{\ \ \ \ \ \ \ \ \ \
\ \ \ \ \ \ \ \ \ \ \ \ \ \ \ \ \ \ } \\ 
\frac{1}{3}\lambda _{2}=\mathcal{L}\lambda _{1}\lambda _{2}\text{ \ \ \ \ \
\ \ \ \ \ \ \ \ \ \ \ \ \ \ \ \ \ \ } \\ 
\frac{1}{3}\left( 2\lambda _{1}+\lambda _{2}\right) =\mathcal{L}\lambda
_{2}^{2}\ \ \text{\ }.\text{ \ \ \ \ \ \ \ \ \ \ \ }%
\end{array}%
\right.
\end{equation*}%
An infinite number of solutions are possible, and correspond to 
\begin{equation*}
\left( \lambda _{1},\lambda _{2}\right) =\frac{1}{3\mathcal{L}_{1}}\left(
1,-1\right) \text{, \ \ \ \ \ \ \ \ \ \ and \ \ \ \ \ }\left( \lambda
_{1},\lambda _{2}\right) =\frac{1}{3\mathcal{L}_{2}}\left( 1,2\right) ,
\end{equation*}%
since there is no need for the proportionality constant to have the same
value in the two cases, i.e. $\xi _{2}$ can differ from $\xi _{1}$. As a
consequence, we can write the two new $h_{\mathrm{new}}(r)$ as 
\begin{equation}
\left\{ 
\begin{array}{c}
h_{1}\left( r\right) =\left( 3\mathcal{L}_{1}\right) ^{-1}\left[ h_{\Delta
}(r)-h_{D}^{0}(r)\right] \text{ \ } \\ 
h_{2}\left( r\right) =\left( 3\mathcal{L}_{2}\right) ^{-1}\left[ h_{\Delta
}(r)+2h_{D}^{0}(r)\right] ,%
\end{array}%
\right.  \label{oz13a}
\end{equation}%
while similar expressions hold for $c_{1}$ and $c_{2}$. From Eqs. (\ref%
{oz12b}) it follows that: $\ h_{1}(r)=K/\mathcal{L}_{1}$ \ and $h_{2}(r)=-2K/%
\mathcal{L}_{2}$\ \ \ for \ $0<r<\sigma .$

In Ref. I Wertheim chose $\mathcal{L}_{1}=-K$ and $\mathcal{L}_{2}=2K$ \cite%
{Wertheim71}, which leads to%
\begin{equation}
\left\{ 
\begin{array}{c}
F_{1}\left( r\right) =\frac{1}{3K}\left[ F_{D}^{0}(r)-F_{\Delta }(r)\right] 
\text{ } \\ 
F_{2}\left( r\right) =\frac{1}{3K}\left[ F_{D}^{0}(r)+\frac{1}{2}F_{\Delta
}(r)\right]%
\end{array}%
\right. \text{\ \ \ \ }\left( F=c,h\right) ,  \label{oz13}
\end{equation}

\begin{equation}
\left\{ 
\begin{array}{c}
\rho _{1}=-K\rho \text{ } \\ 
\rho _{2}=2K\rho ,%
\end{array}%
\right. \qquad \qquad \left\{ 
\begin{array}{c}
h_{1}(r)=-1 \\ 
h_{2}(r)=-1%
\end{array}%
\right. \text{ \ \ \ \ \ for \ }0<r<\sigma
\end{equation}%
(in Ref. I, $F_{1}$ and $F_{2}$ were denoted as $F_{-}$ and $F_{+}$,
respectively). Clearly, Wertheim's choice has the advantage of providing,
for all the three `hypothetical' fluids, core conditions of the typical HS
form: $h_{m}(r)=-1$ for \ $0<r<\sigma $ ($m=0,1,2$). The cost to pay is the
introduction of `modified densities' for the `auxiliary' fluids $1$ and $2$
(the negative sign of $\rho _{1}$ poses no special difficulty).

On the other hand, it is would be equally proposable the choice $\mathcal{L}%
_{1}=\mathcal{L}_{2}=1$, which leads to%
\begin{equation*}
\left\{ 
\begin{array}{c}
F_{1}\left( r\right) =\frac{1}{3}\ \left[ F_{\Delta }(r)-F_{D}^{0}(r)\right] 
\text{ \ \ } \\ 
F_{2}\left( r\right) =\frac{1}{3}\ \left[ F_{\Delta }(r)+2F_{D}^{0}(r)\right]
\text{ }%
\end{array}%
\right. \text{\ \ \ \ }\left( F=c,h\right) ,
\end{equation*}%
\begin{equation*}
\left\{ 
\begin{array}{c}
\rho _{1}=\rho \text{ } \\ 
\rho _{2}=\rho ,%
\end{array}%
\right. \qquad \qquad \left\{ 
\begin{array}{c}
h_{1}(r)=K\text{ \ \ } \\ 
h_{2}(r)=-2K%
\end{array}%
\right. \text{ \ \ \ \ \ for \ }0<r<\sigma .
\end{equation*}%
The advantage of this second possibility would be that all the three
`hypothetical' fluids have the same \textit{real} density, while the cost is
represented by the less usual core conditions, which however pose no
particular difficulty.

\bigskip

\section{Equations for the unknown parameters}

Three \textit{quadratic} equations for the $\Lambda _{m}$\ $^{\prime }s$ $%
(m=0,1,2)$ can be obtained from Eqs. ($\ref{gpy3}$)-($\ref{gpy4}$), after
deriving from Eq. ($\ref{so7}$) the following expressions for the PY-OL 
\textit{contact values}

\begin{equation}
h_{m,\text{\textrm{reg}}}\left( \sigma ^{+}\right) =\ h_{\sigma }^{\mathrm{HS%
}}\left( \eta _{m}\right) -\frac{12\eta _{m}}{1-\eta _{m}}\ \Lambda
_{m}+12\eta _{m}\ \Lambda _{m}^{2},  \label{b1}
\end{equation}%
where%
\begin{equation}
\begin{array}{c}
\text{\ \ \ \ \ \ \ \ \ \ \ \ \ \ \ \ \ }h_{\sigma }^{\mathrm{HS}}\left(
x\right) =y_{\sigma }^{\mathrm{HS}}\left( x\right) -1\text{ \ \ \ \ \ \ \ \
\ \ \ \ \ \ \ \ \ \ \ \ } \\ 
y_{\sigma }^{\mathrm{HS}}\left( x\right) =\left( 1+\frac{1}{2}x\right)
\left( 1-x\right) ^{-2}.%
\end{array}
\label{b2}
\end{equation}%
Substituting Eq. (\ref{b1}) into the expressions for $\Lambda _{m}$ given by
Eqs. ($\ref{gpy3}$), we get:

i) for $\Lambda _{0}$, the same PY equation found by Baxter for isotropic
SHS \cite{Baxter68,Baxter71} 
\begin{equation}
12\eta t\ \Lambda _{0}^{2}-\left( 1+\frac{12\eta }{1-\eta }t\right) \Lambda
_{0}+y_{\sigma }^{\mathrm{HS}}(\eta )t=0.  \label{App_b5}
\end{equation}%
Only the smaller of the two real solutions (when they exist) is physically
significant \cite{Baxter68,Baxter71}, and reads

\begin{equation}
\Lambda _{0}=\frac{y_{\sigma }^{\mathrm{HS}}(\eta )\ t}{\frac{1}{2}\left[ 1+%
\frac{12\eta }{1-\eta }t+\sqrt{\left( 1+\frac{12\eta }{1-\eta }t\right)
^{2}-48\eta \ y_{\sigma }^{\mathrm{HS}}(\eta )\ t^{2}}\right] }.  \label{b6}
\end{equation}

ii) For $\Lambda _{1}$ and $\Lambda _{2}$, the equations \ \ \ \ 
\begin{equation}
12\eta _{m}t\ \Lambda _{m}^{2}-\left( 1+\frac{12\eta _{m}}{1-\eta _{m}}%
t\right) \Lambda _{m}+h_{\sigma }^{\mathrm{HS}}(\eta _{m})t=-\mathcal{P}%
\text{\quad ~~}(m=1,2).  \label{b7}
\end{equation}%
It is remarkable that the right-hand member of these equations does not
depend on the index $m$. This fact means that $\Lambda _{2}$ obeys exactly
the same equation as $\Lambda _{1}$, but with $\eta _{2}$ replacing $\eta
_{1}$; as will be confirmed later, such a property implies that, if one
writes $\Lambda _{1}=\Lambda _{1}\left( \eta _{1},\eta _{2},t,\alpha \right) 
$, then $\Lambda _{2}$ must have the same functional form with $\eta _{2}$
interchanged with $\eta _{1}$, i.e. $\Lambda _{2}\left( \eta _{1},\eta
_{2},t,\alpha \right) =\Lambda _{1}\left( \eta _{2},\eta _{1},t,\alpha
\right) .$

Now the system of equations for $\Lambda _{1}$, $\Lambda _{2}$ and $K$ must
be completed by a further relationship, which can be obtained from the sum
rule, Eq. (\ref{f4b}). Taking into account that $c_{D}^{0}=K\left(
2c_{2}+c_{1}\right) $, and multiplying Eq. (\ref{f4b}) by $4\pi \rho $ yields

\begin{equation}
4\pi \rho _{2}\ \int_{0}^{\infty }\ c_{2}(x)\ x^{2}\ dx=4\pi \rho
_{1}\int_{0}^{\infty }c_{1}(x)\ x^{2}\ dx,  \label{k4b}
\end{equation}%
On the other hand, putting $k=0$ into Eq. ($\ref{bf1}$) gives%
\begin{equation*}
1-\rho _{m}\widetilde{c}_{m}\left( k=0\right) =1-4\pi \rho
_{m}\int_{0}^{\infty }\ c_{m}(r)\ r^{2}\ dr=Q_{m}^{2}(k=0)=\ a_{m}^{2}\ ,
\end{equation*}%
since $Q_{m}(k=0)\equiv a_{m}$ (as shown by the first of Eqs. ($\ref{ie4}$)
). Then Eq. ($\ref{k4b}$) becomes $\ a_{2}^{2}=\ a_{1}^{2},$ which splits
into two equations: $\ \ \ a_{2}=a_{1},$ and $\ a_{2}=-a_{1}$. From the
expression for $a_{m}$, one can easily realize that the second equation does
not satisfy the $t\rightarrow 0$ limit, whereas the first one, $a_{2}\ =\
a_{1}$ (or, equivalently, $a^{\mathrm{isoSHS}}(\eta _{2},\Lambda _{2})=a^{%
\mathrm{isoSHS}}(\eta _{1},\Lambda _{1})\ $), leads to the following \textit{%
linear }relationship between $\Lambda _{1}$ and $\Lambda _{2}$

\begin{equation}
\frac{12\eta _{2}\ \Lambda _{2}\ }{1-\eta _{2}}-\frac{12\eta _{1}\ \Lambda
_{1}\ }{1-\eta _{1}}=\ a^{\mathrm{HS}}(\eta _{2})-\ a^{\mathrm{HS}}(\eta
_{1}).  \label{b8}
\end{equation}

Note that the two Eqs. (\ref{b7}) are coupled (since $K_{\mathrm{reg}%
}/K=1-\Lambda _{D}/K=1-(2\Lambda _{2}+\Lambda _{1})\ $), but with the help
of Eq. (\ref{b8}) they could be easily decoupled. However, since the
right-hand members of Eqs. (\ref{b7}) coincide, we can get a new
relationship by equating their first members, and exploiting Eq. (\ref{b8}).
So we arrive at the following equations for the three unknowns $\Lambda _{1}$%
, $\Lambda _{2}$ and $K$:%
\begin{equation}
\left\{ 
\begin{array}{c}
12\eta _{2}t\ \Lambda _{2}^{2}-\Lambda _{2}+b^{\mathrm{HS}}(\eta
_{2})t=12\eta _{1}t\ \Lambda _{1}^{2}-\Lambda _{1}+b^{\mathrm{HS}}(\eta
_{1})t\text{ \ \ } \\ 
\frac{12\eta _{2}\ \Lambda _{2}\ }{1-\eta _{2}}-\frac{12\eta _{1}\ \Lambda
_{1}\ }{1-\eta _{1}}=\frac{\eta _{2}\left( 4-\eta _{2}\right) \ }{\left(
1-\eta _{2}\right) ^{2}}-\frac{\eta _{1}\left( 4-\eta _{1}\right) \ }{\left(
1-\eta _{1}\right) ^{2}}\ \ \ \ \ \ \ \ \ \ \ \ \ \ \ \ \ \ \  \\ 
12\eta _{1}t\ \Lambda _{1}^{2}-\left( 1+\frac{12\eta _{1}}{1-\eta _{1}}%
t\right) \Lambda _{1}+h_{\sigma }^{\mathrm{HS}}(\eta _{1})t=-\mathcal{P}.%
\text{ \ \ \ \ \ \ \ \ \ \ \ \ \ \ \ }%
\end{array}%
\right.  \label{s2}
\end{equation}%
The first two equations form a closed system for $\Lambda _{1}$ and $\Lambda
_{2}$. The second one suggests that we can assume

\begin{equation*}
\frac{12\eta _{m}\ \Lambda _{m}\ }{1-\eta _{m}}=\frac{\eta _{m}\left( 4-\eta
_{m}\right) \ }{\left( 1-\eta _{m}\right) ^{2}}+W,
\end{equation*}%
or, equivalently,%
\begin{equation}
\ \Lambda _{m}\ =\frac{1}{3}+\frac{\eta _{m}}{4\left( 1-\eta _{m}\right) }+%
\frac{1-\eta _{m}}{12\eta _{m}}\ W\text{\ \ \ \ \ \ \ }(m=1,2),\text{ }
\label{s3}
\end{equation}%
where $W=W(\eta _{1},\eta _{2},t)$ is an unknown function, which must be
proportional to $\eta _{1}\eta _{2}$. In fact, Eqs. (\ref{b7}) require that%
\begin{equation}
\lim_{\eta \rightarrow 0}\Lambda _{1}=\lim_{\eta \rightarrow 0}\Lambda _{2}=%
\frac{1}{3}\ ,  \label{s3b}
\end{equation}%
since, from Eq. (\ref{gpy4}), one has $\lim_{\eta \rightarrow 0}\mathcal{P}=%
\frac{1}{3}$ ( $\lim_{\eta \rightarrow 0}K_{\mathrm{reg}}=0$). If $\Lambda
_{1}$ and $\Lambda _{2}$ in the first of \ Eqs. (\ref{s2}) are replaced with
the new expressions (\ref{s3}), then one gets a quadratic equation for $W$:

\begin{equation}
\allowbreak \allowbreak \left( 1-\eta _{1}\eta _{2}\right) t\ W^{2}-\left(
1-2\eta _{1}\eta _{2}t\right) W+\allowbreak \frac{3\eta _{1}\eta _{2}}{%
\left( 1-\eta _{1}\right) \left( 1-\eta _{2}\right) }M\allowbreak
\allowbreak =0\   \label{s4a}
\end{equation}%
with%
\begin{eqnarray}
M\allowbreak &=&\allowbreak 1+\left[ \frac{1+\allowbreak 2\left( \eta
_{1}+\eta _{2}\right) -5\eta _{1}\eta _{2}}{\left( 1-\eta _{1}\right) \left(
1-\eta _{2}\right) }-\frac{1}{3}\left( 1-\eta _{1}\right) \left( 1-\eta
_{2}\right) \right] t  \notag  \label{ssss} \\
&=&1+\left[ \frac{1+\allowbreak 2x+10x^{2}}{\left( 1+x\right) \left(
1-2x\right) }-\frac{1}{3}\left( 1+x\right) \left( 1-2x\right) \right] t,
\label{s5}
\end{eqnarray}%
where we have put $\eta _{1}=-x$, $\eta _{2}=2x$ \ ( $x\equiv K\eta $ ). The
acceptable solution is%
\begin{eqnarray}
W &=&\frac{1-2\eta _{1}\eta _{2}t}{2\left( 1-\eta _{1}\eta _{2}\right) t}%
\left( 1-\sqrt{\mathcal{D}}\right)  \notag \\
&=&\frac{3\eta _{1}\eta _{2}}{\left( 1-\eta _{1}\right) \left( 1-\eta
_{2}\right) }W_{0}=-\ \frac{6x^{2}}{\left( 1+x\right) \left( 1-2x\right) }%
W_{0}
\end{eqnarray}%
with%
\begin{equation}
W_{0}=\frac{\ \allowbreak M}{\frac{1}{2}\left( 1-2\eta _{1}\eta _{2}t\right)
\left( 1+\sqrt{\mathcal{D}}\right) }=\frac{\ \allowbreak M}{\frac{1}{2}%
\left( 1+4x^{2}t\right) \left( 1+\sqrt{\mathcal{D}}\right) },
\end{equation}

\begin{eqnarray}
\mathcal{D} &=&1-\allowbreak \frac{12\eta _{1}\eta _{2}\left( 1-\eta
_{1}\eta _{2}\right) }{\left( 1-\eta _{1}\right) \left( 1-\eta _{2}\right)
\left( 1-2\eta _{1}\eta _{2}t\right) ^{2}}\ Mt  \notag \\
&=&1+\ \allowbreak \frac{24x^{2}\left( 1+2x^{2}\right) }{\left( 1+x\right)
\left( 1-2x\right) \left( 1+4x^{2}t\right) ^{2}}\ Mt.  \label{s6}
\end{eqnarray}%
Note that $\lim_{\eta \rightarrow 0}W_{0}=\lim_{\eta \rightarrow
0}M=1+\left( 2/3\right) t$.

\bigskip The functions $W$, $W_{0}$, $D$ and $M$ are symmetrical with
respect to the exchange of $\eta _{1}$ and $\eta _{2}$; in particular, $%
W(\eta _{2},\eta _{1},t)=W(\eta _{1},\eta _{2},t)$, and this property
implies that%
\begin{equation}
\Lambda _{2}\left( \eta _{1},\eta _{2},t\right) =\Lambda _{1}\left( \eta
_{2},\eta _{1},t\right) ,  \label{s7}
\end{equation}%
confirming our previous guess.

Moreover, if we put%
\begin{equation}
W_{0}=1+W_{0}^{\mathrm{ex}},
\end{equation}%
then%
\begin{equation}
\Lambda _{m}=\Lambda +\Lambda _{m}^{\mathrm{ex}}\ ,  \label{r1}
\end{equation}%
with%
\begin{equation}
\Lambda =\frac{1}{3}+\frac{1}{4}\left( \frac{\eta _{1}}{1-\eta _{1}}+\frac{%
\eta _{2}}{1-\eta _{2}}\right) =\frac{1}{3}+\allowbreak \frac{x(1+4x)}{%
4\left( 1+x\right) \left( 1-2x\right) },  \label{r2}
\end{equation}%
\begin{equation}
\Lambda _{1}^{\mathrm{ex}}=\frac{\eta _{2}}{4\left( 1-\eta _{2}\right) }%
W_{0}^{\mathrm{ex}},\qquad \Lambda _{2}^{\mathrm{ex}}=\frac{\eta _{1}}{%
4\left( 1-\eta _{1}\right) }W_{0}^{\mathrm{ex}}.  \label{r3}
\end{equation}%
Here, both $\Lambda $ and $W_{0}^{\mathrm{ex}}$ are symmetric with respect
to $\eta _{1}$ and $\eta _{2}$, whereas $\Lambda _{m}^{\mathrm{ex}}$
represents the asymmetric part of $\Lambda _{m}$.

Note that the knowledge of $\Lambda _{1}$ and $\Lambda _{2}$ allows to
calculate $\Lambda _{\Delta }$ and $\Lambda _{D}$ immediately. In fact, Eqs.
(\ref{c10})\ lead to%
\begin{equation}
\left\{ 
\begin{array}{c}
\Lambda _{\Delta }=2K\left( \Lambda _{2}-\Lambda _{1}\right) =-K\ \frac{3x}{%
2\left( 1+x\right) \left( 1-2x\right) }\ W_{0}^{\mathrm{ex}}\text{\ \ \ \ \
\ \ \ \ \ \ \ \ \ \ \ \ \ \ \ \ \ \ \ \ \ } \\ 
\Lambda _{D}=K\left( 2\Lambda _{2}+\Lambda _{1}\right) =K\left\{ \allowbreak
1+\allowbreak \frac{3x}{4\left( 1+x\right) \left( 1-2x\right) }\left[
1+2x\left( 2+W_{0}^{\mathrm{ex}}\right) \right] \right\} \text{.\ }%
\end{array}%
\right.  \label{s8}
\end{equation}

Now we must find an equation for $K$. We can regard the third of Eqs. (\ref%
{s2}) as the required relationship. However, in order to derive a more
symmetric expression, we prefer to start from Eqs. (\ref{b7}), rewritten as 
\begin{equation}
\left\{ 
\begin{array}{c}
12\eta _{1}t\ \Lambda _{1}^{2}-\left( 1+\frac{12\eta _{1}}{1-\eta _{1}}%
t\right) \Lambda _{1}+h_{\sigma }^{\mathrm{HS}}(\eta _{1})t+\frac{K_{\mathrm{%
reg}}}{K}t+\frac{1}{3}\frac{\alpha t}{K}\ y_{0}^{\mathrm{PY}}(\sigma )=0%
\text{ } \\ 
12\eta _{2}t\ \Lambda _{2}^{2}-\left( 1+\frac{12\eta _{2}}{1-\eta _{2}}%
t\right) \Lambda _{2}+h_{\sigma }^{\mathrm{HS}}(\eta _{2})t+\frac{K_{\mathrm{%
reg}}}{K}t+\frac{1}{3}\frac{\alpha t}{K}\ y_{0}^{\mathrm{PY}}(\sigma )=0,%
\end{array}%
\right.  \label{s8b}
\end{equation}%
and we get%
\begin{equation}
K=\alpha t\ \mathcal{K},\text{ }\ \ \ \ \ \text{with \ \ \ }\ \mathcal{K}%
\text{ }=\frac{y_{0}^{\mathrm{PY}}(\sigma )}{Z(\eta _{1},\eta _{2},t)},
\label{s9}
\end{equation}%
\begin{equation}
Z=\frac{3}{2}\left( \Lambda _{1}+\Lambda _{2}\right) -3\left\{ \frac{1}{2}%
\sum_{m=1}^{2}\ \left[ 12\eta _{m}\ \Lambda _{m}^{2}-\frac{12\eta
_{m}\Lambda _{m}}{1-\eta _{m}}+h_{\sigma }^{\mathrm{HS}}(\eta _{m})\right] +%
\frac{K_{\mathrm{reg}}}{K}\right\} t  \label{s10}
\end{equation}%
and $\lim_{\eta \rightarrow 0}Z(\eta _{1},\eta _{2},t)=1$. Replacing the
found expressions for $\Lambda _{1},$ $\Lambda _{2}$ and $\Lambda _{D}$ into
Eq. (\ref{s9}) yields an equation for $K$ that we have solved numerically,
although some further analytic simplifications are probably possible.

\bigskip

\bigskip

\bigskip 

\bigskip

\newpage 
\begin{figure}[tbph]
\begin{center}
\includegraphics[width=10cm]{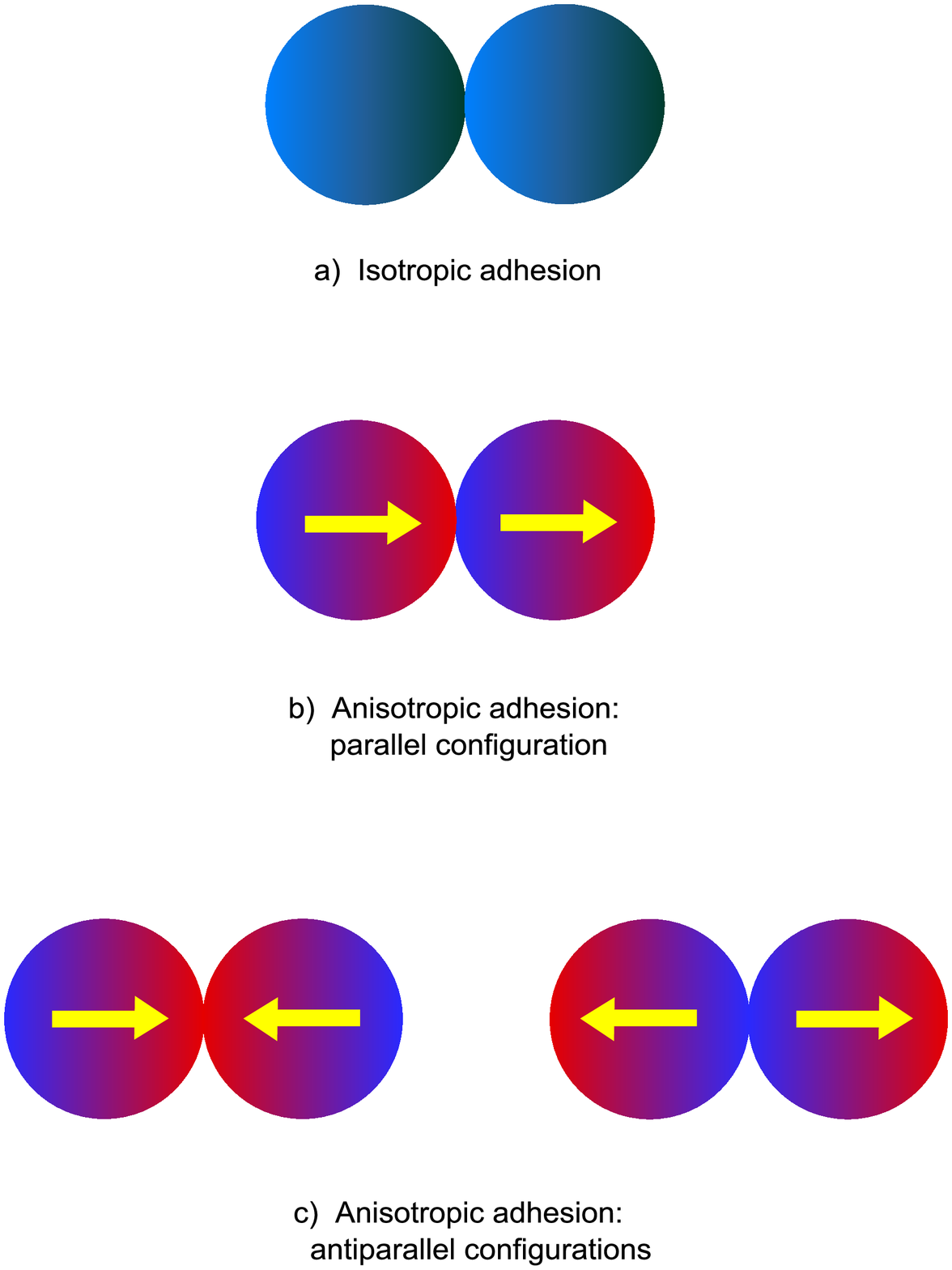}
\end{center}
\caption{(Color online) Illustration of the dipolar-like adhesion. In the
top panel a) the adhesion is isotropic, with $\protect\epsilon (1,2) = 1$.
In the two other cases the adhesion is anisotropic and: i) stronger and
maximum in the head-to-tail parallel configuration b), where $\protect%
\epsilon (1,2) = 1 + 2 \protect\alpha$; ii) weaker and minimum in the two
antiparallel configurations c) ( head-to-head and tail-to-tail orientations,
both with $\protect\epsilon (1,2) = 1 - 2 \protect\alpha$). }
\label{fig1}
\end{figure}

%
%
\begin{figure}[tbph]
\begin{center}
\includegraphics[width=8cm]{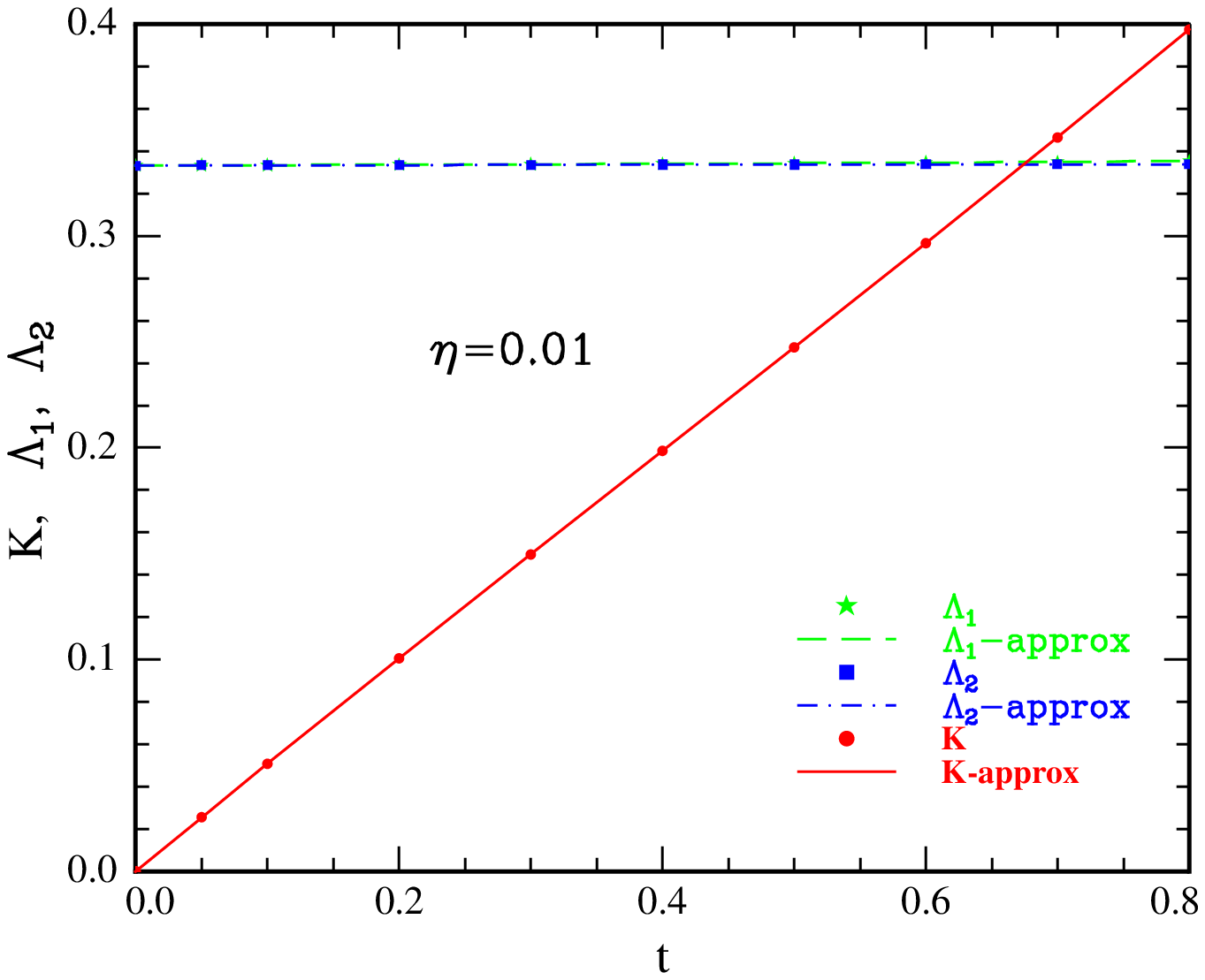}\\[0pt]
\includegraphics[width=8cm]{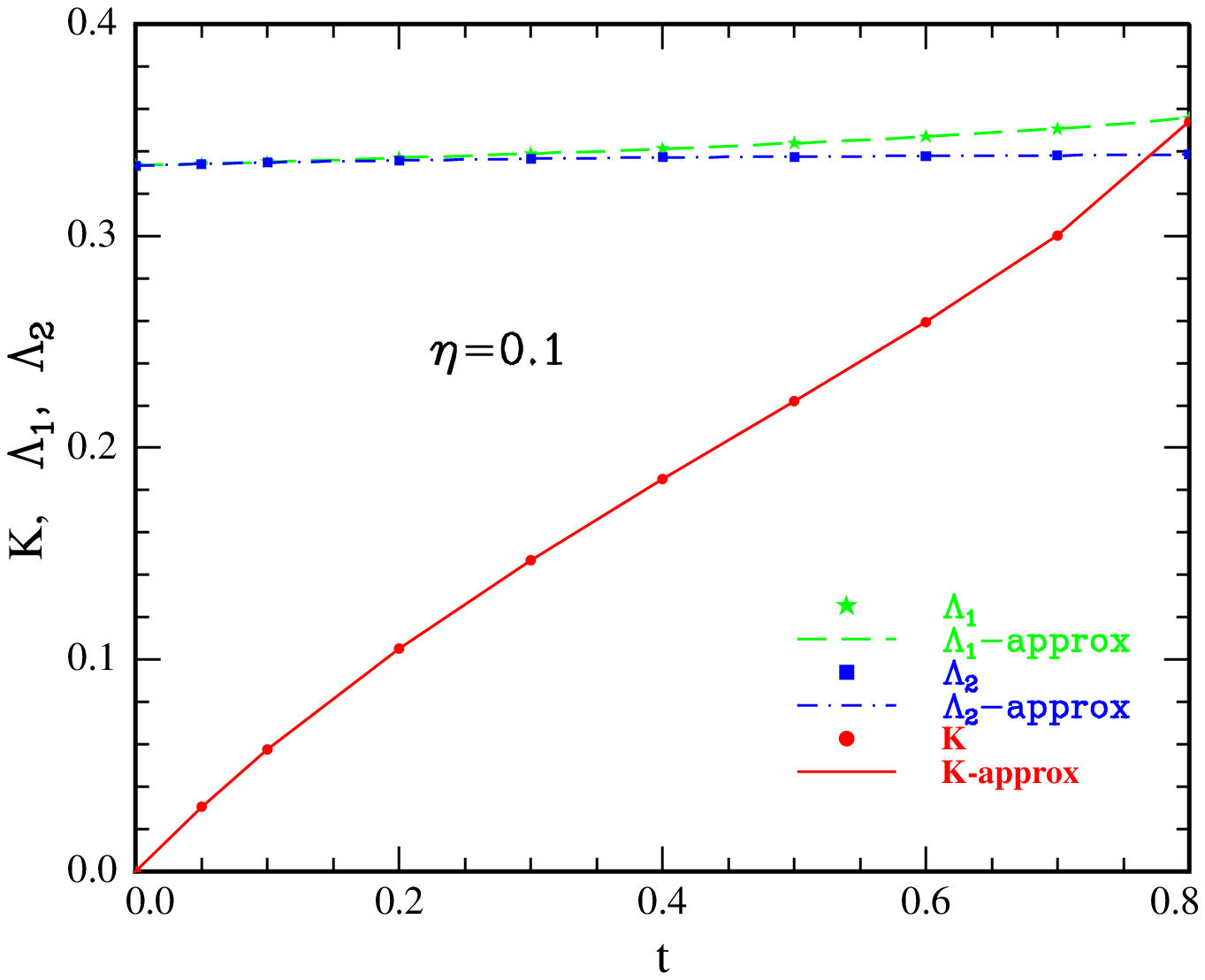}\\[0pt]
\end{center}
\caption{(Color online) Comparison between exact numerical and approximate
analytical results for the parameters $K$, $\Lambda _{1},$ and $\Lambda _{2}$
as a function of $t$, for anisotropy degree $\protect\alpha = 1/2$ and two
values of the packing fraction: $\protect\eta = 0.01$ (top panel) and $%
\protect\eta = 0.1$ (bottom panel). }
\label{fig2}
\end{figure}
%
%
\begin{figure}[tbph]
\begin{center}
\includegraphics[width=8cm]{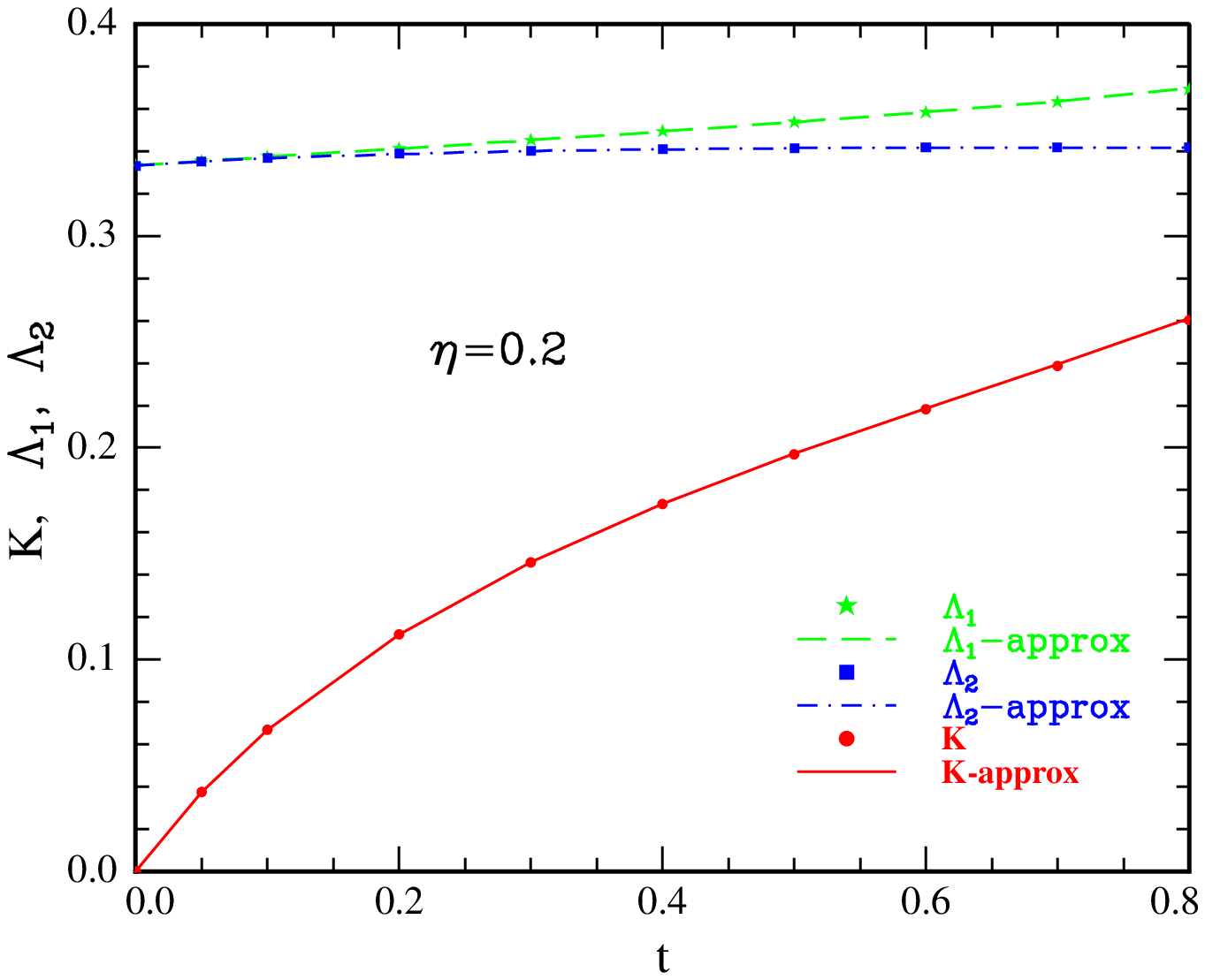}\\[0pt]
\includegraphics[width=8cm]{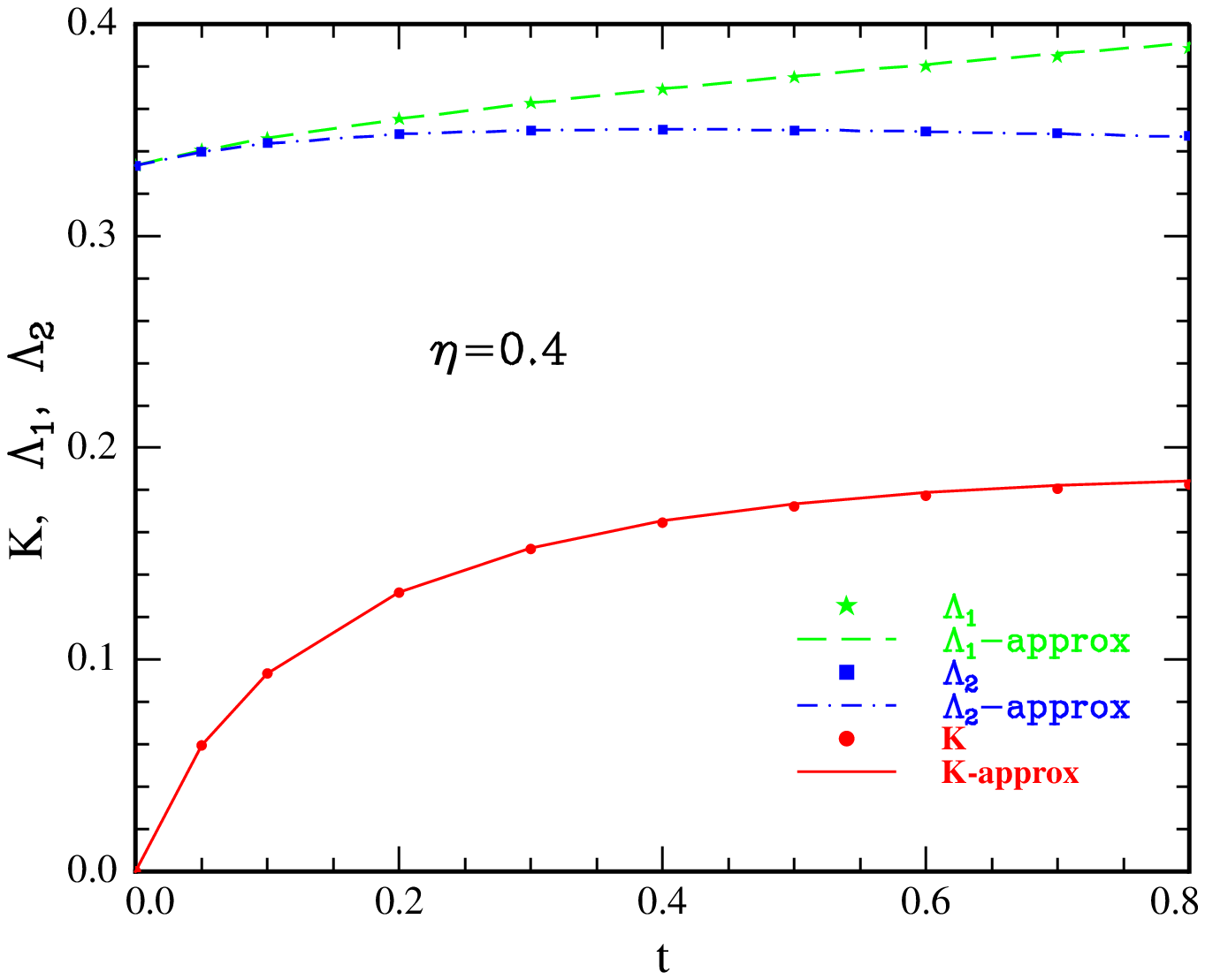}\\[0pt]
\end{center}
\caption{(Color online) Same as in Figure 2, but for $\protect\eta = 0.2$
(top panel) and $\protect\eta = 0.4$ (bottom panel). }
\label{fig3}
\end{figure}

\begin{figure}[tbph]
\begin{center}
\includegraphics[width=10cm]{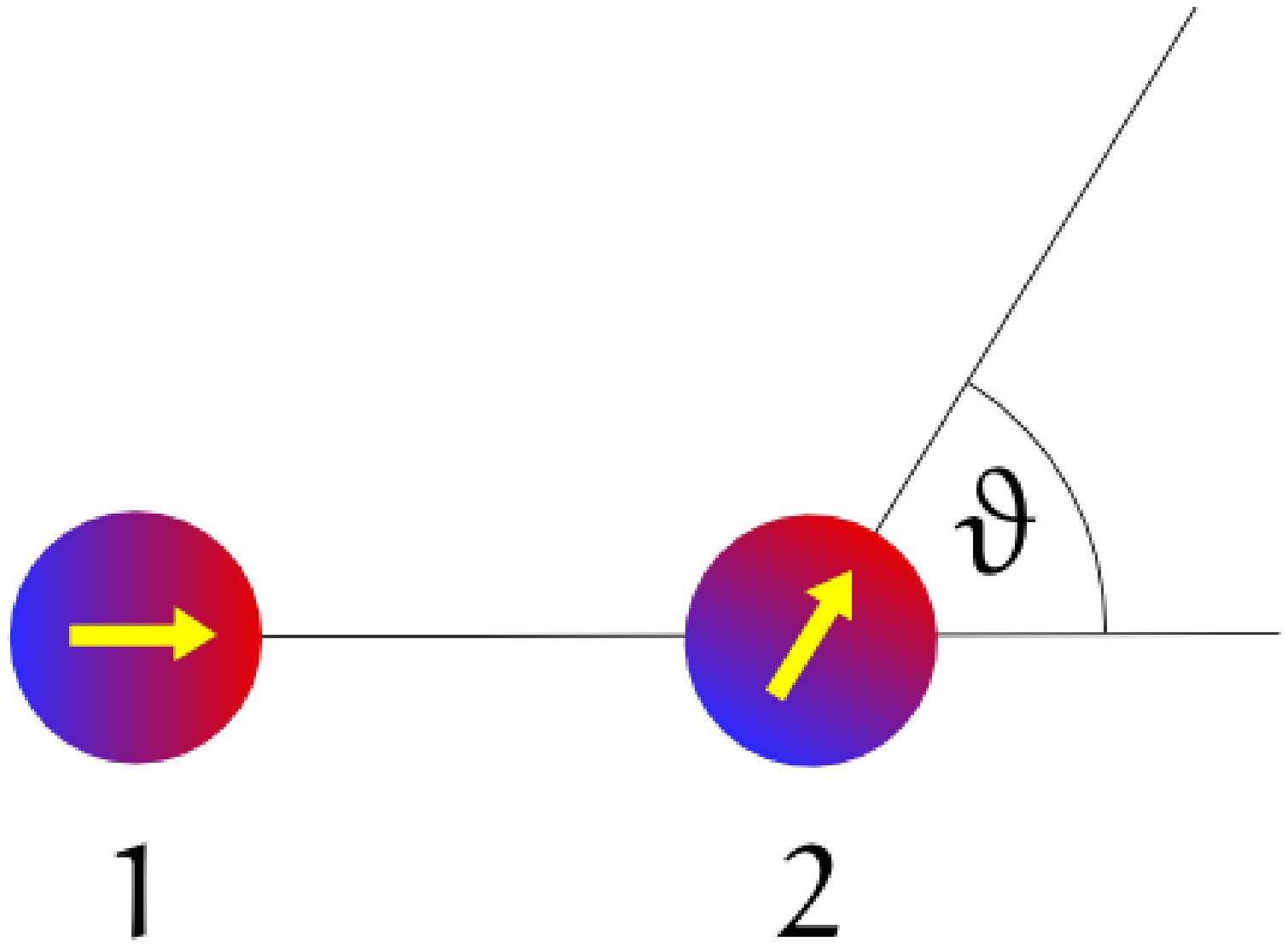}
\end{center}
\caption{(Color online) Illustration of the simple configuration discussed
in the text and chosen to define some radial sections through the
multidimensional plot of $g(1,2)$. }
\label{fig4}
\end{figure}
%
%
\begin{figure}[tbph]
\begin{center}
\includegraphics[width=8cm]{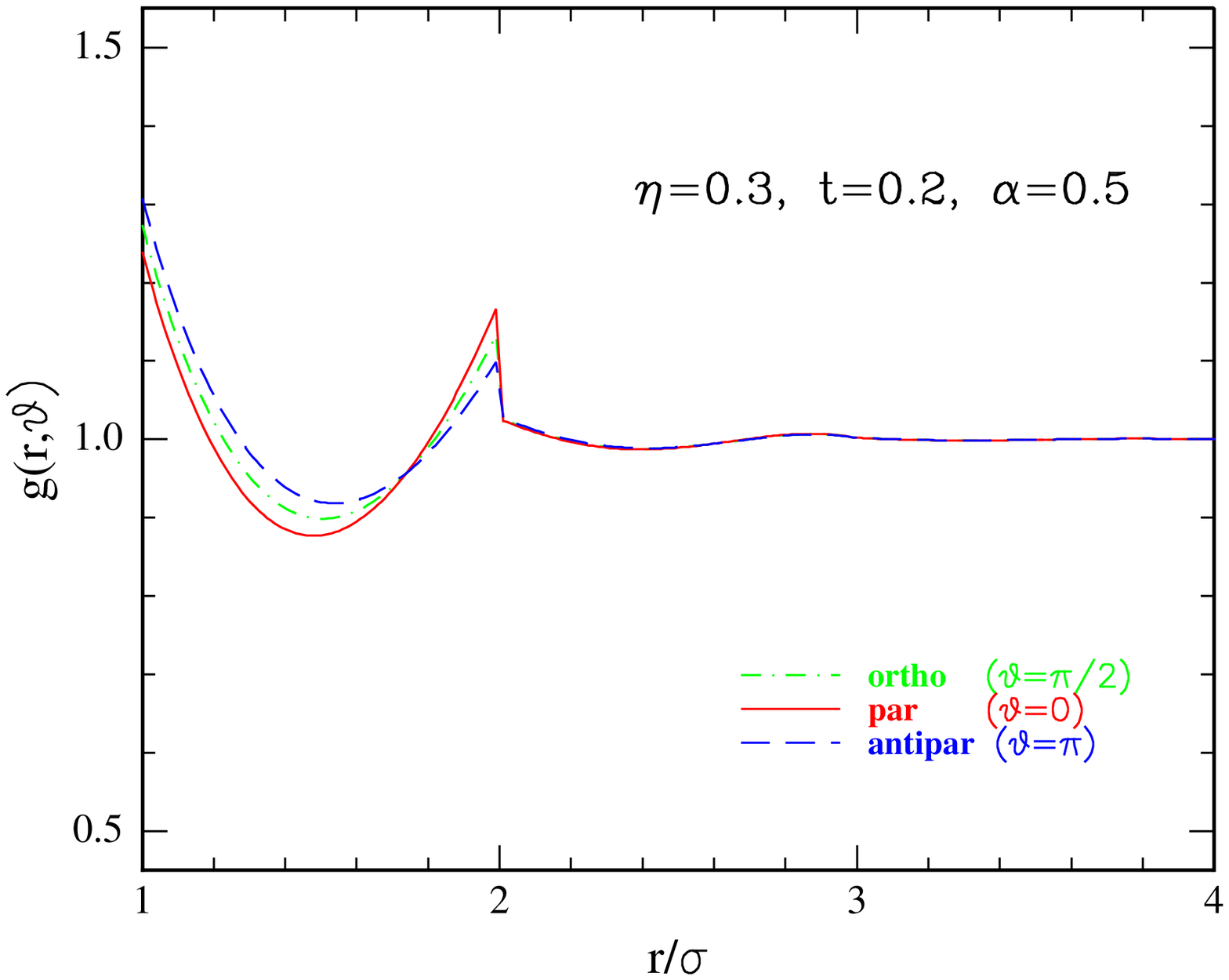}\\[0pt]
\includegraphics[width=8cm]{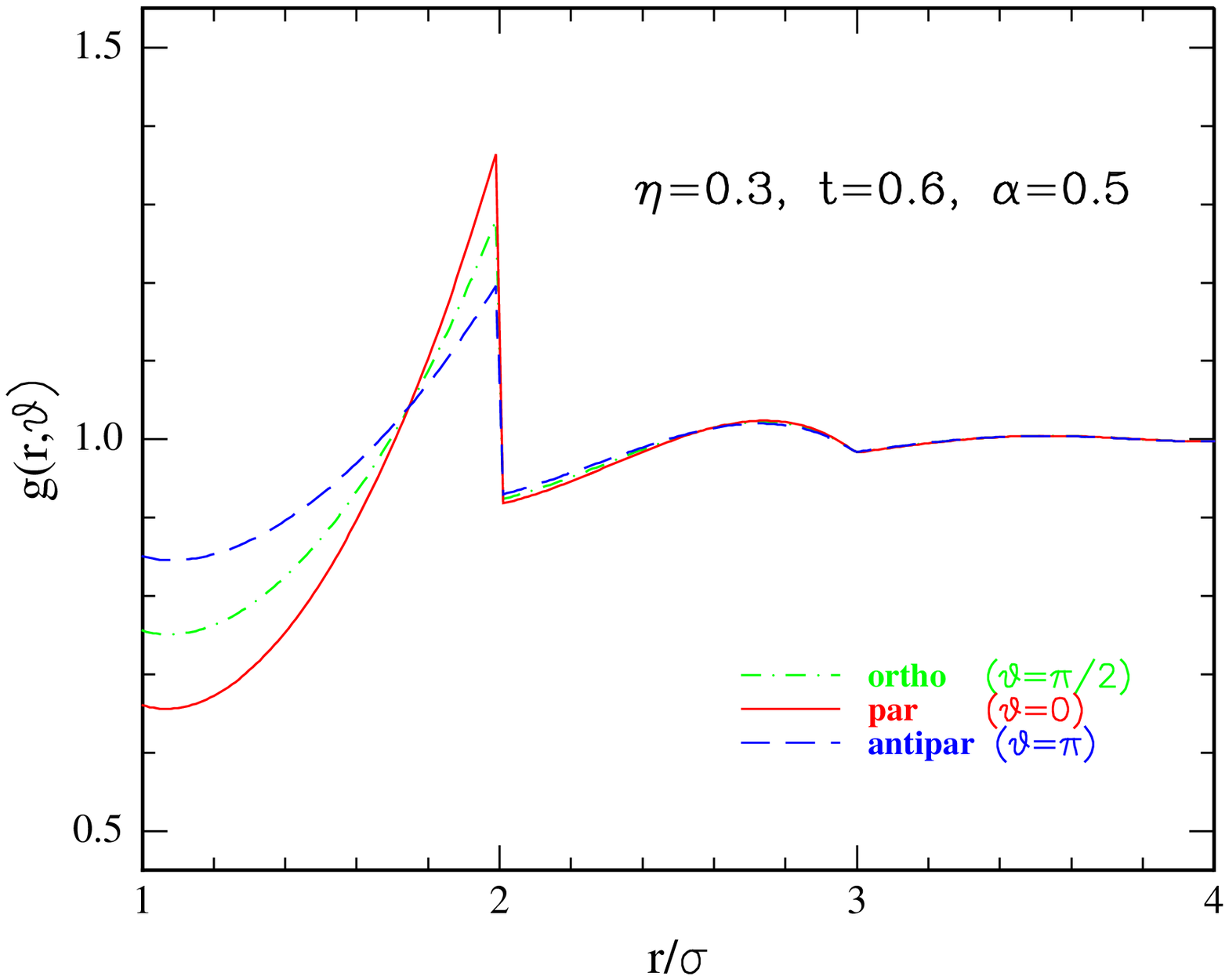}\\[0pt]
\end{center}
\caption{(Color online) Sections through $g(1,2)$, with particles 1 and 2 in
the configuration shown by the previous Figure, calculated as a function of $%
r$ for fixed relative orientations: $\protect\theta =0$ (parallel
configuration), $\protect\theta = \protect\pi /2$ (orthogonal
configuration), and $\protect\theta = \protect\pi$ (antiparallel
configurations). }
\label{fig5}
\end{figure}
%
\begin{figure}[tbph]
\begin{center}
\includegraphics[width=8cm]{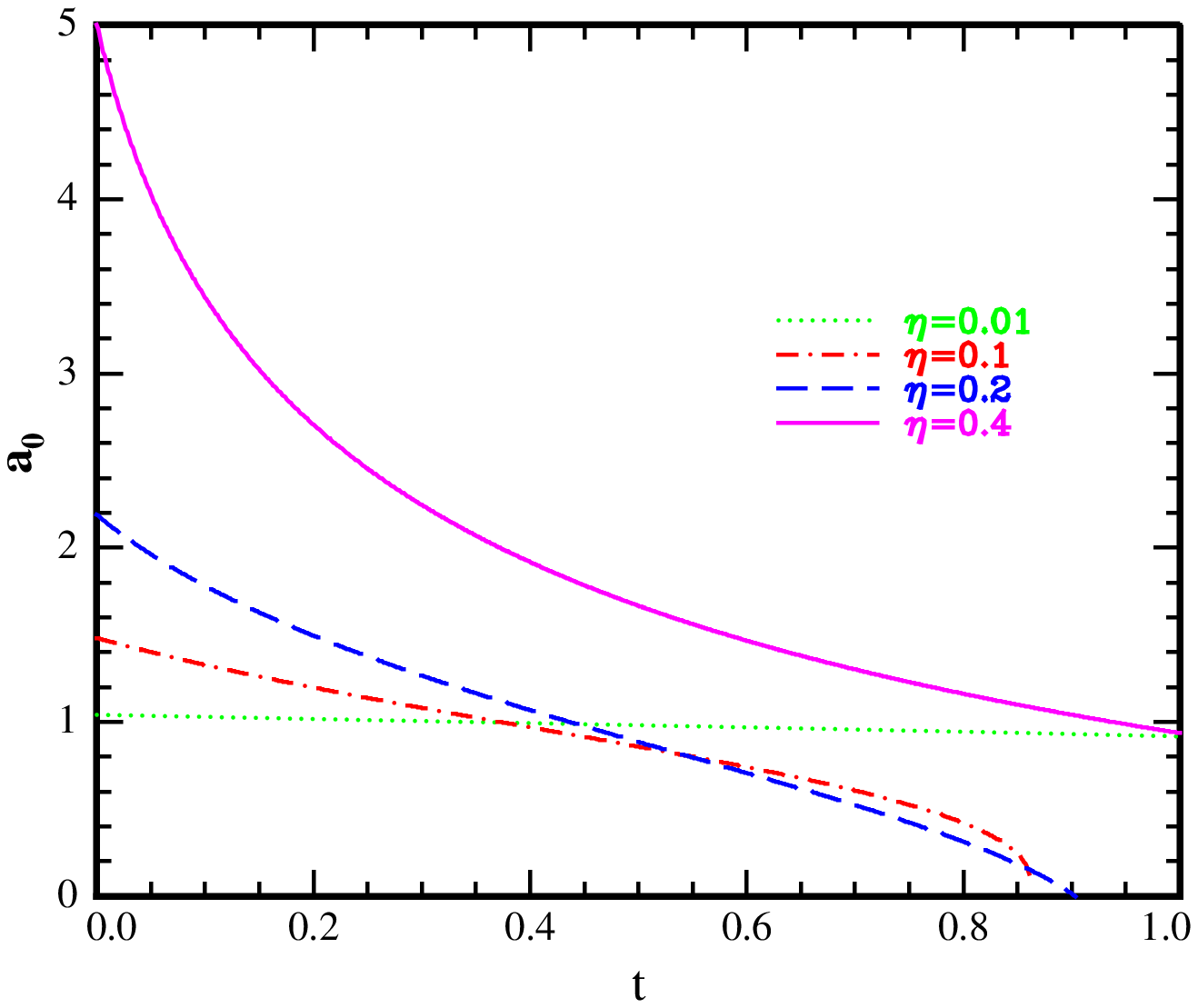}\\[0pt]
\includegraphics[width=8cm]{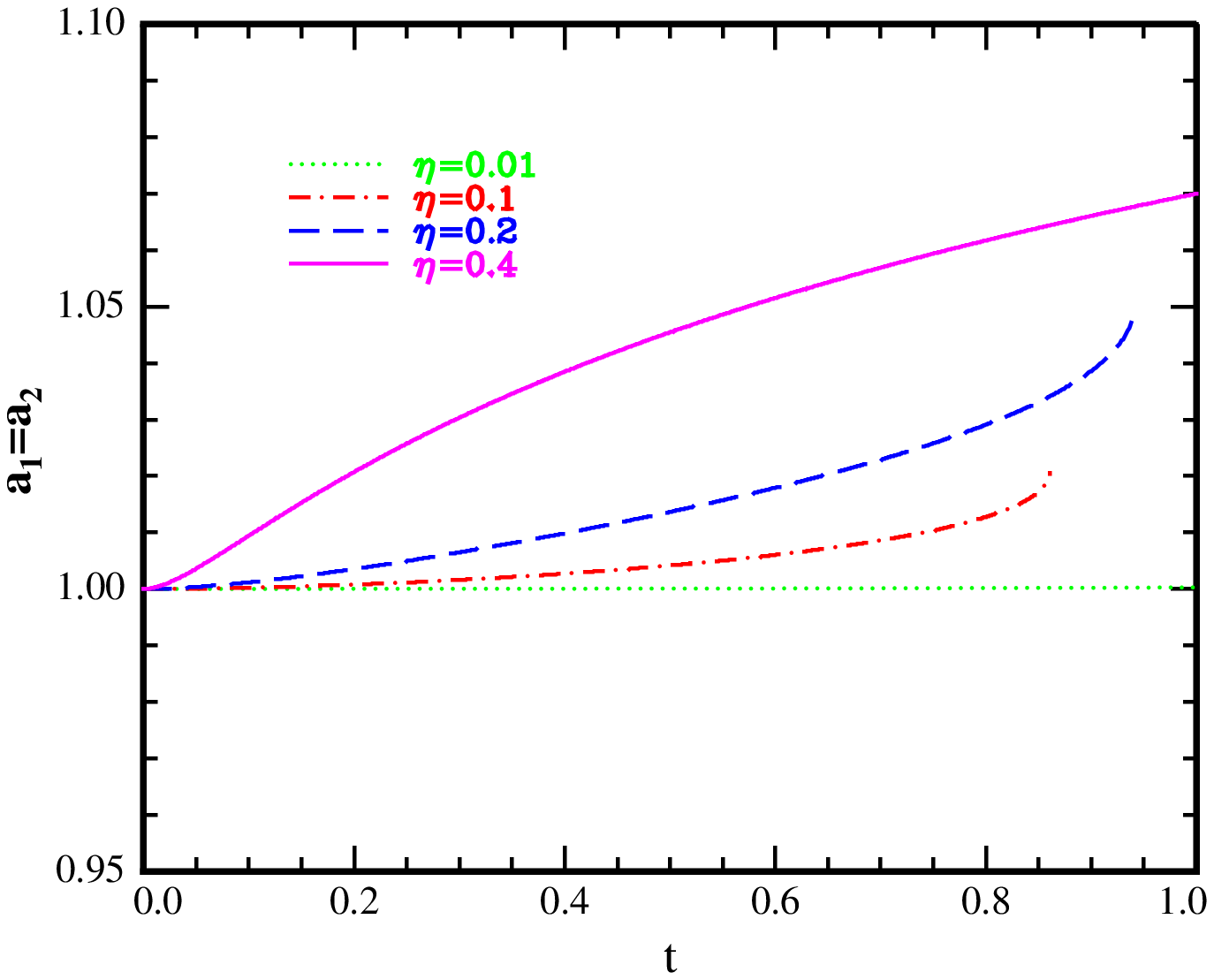}\\[0pt]
\end{center}
\caption{(Color online) Evaluation of quantities $a_0$ (top panel) and $a_1=a_2$ (bottom panel) 
as a function of $t$ for various packing fractions ranging from $\eta=0.01$ to $\eta=0.4$. These
are computed from Eq.(\ref{so2}) with $m=0,1$. Note that for both $\eta=0.1$ and $\eta=0.2$,
$a_0=0$ corresponds to the onset of isotropic instability.}
\label{fig6}
\end{figure}


\begin{thebibliography}{99}
\bibitem{Spinozzi02} F. Spinozzi, D. Gazzillo, A. Giacometti, P. Mariani,
and F. Carsughi, Biophysical Journal \textbf{82}, 2165 (2002).

\bibitem{Giacometti05} A. Giacometti, D. Gazzillo, G. Pastore, and T. Kanti
Das, Phys. Rev. E \textbf{71}, 031108 (2005).

\bibitem{Wertheim71} M. S. Wertheim, J. Chem. Phys. \textbf{55}, 4291 (1971).

\bibitem{Weis93} J. J. Weis and D. Levesque, Phys. Rev Lett. \textbf{71},
2729 (1993).

\bibitem{Leeuwen93} M. E. van Leeuwen and B. Smit, Phys. Rev Lett. \textbf{71%
}, 3991 (1993).

\bibitem{Sear96} R. P. Sear, Phys. Rev Lett. \textbf{76}, 2310 (1996).

\bibitem{Camp00} P.J. Camp, J.C. Shelly and G.N. Patey, Phys. Rev. Lett. 
\textbf{84}, 115 (2000).

\bibitem{Tlusty00} T. Tlusty and S.A. Safran, Science \textbf{290}, 1328
(2000).

\bibitem{Gazzillo06} D. Gazzillo, A. Giacometti, R. Fantoni, and P. Sollich,
Phys. Rev. E \textbf{74}, 051407 (2006).

\bibitem{Fantoni07} R. Fantoni, D. Gazzillo, A. Giacometti, M.A. Miller and
G. Pastore, J. Chem. Phys. \textbf{127}, 234507 (2007).

\bibitem{Baxter68} R. J. Baxter, J. Chem. Phys. \textbf{49}, 2770 (1968).

\bibitem{Baxter71} R. J. Baxter, in \textit{Physical Chemistry, an Advanced
Treatise, }Vol. 8A, ed. by D. Henderson (Academic, New York, 1971) Chap. 4.

\bibitem{Sciortino05} F. Sciortino, P. Tartaglia, and E. Zaccarelli, J.
Phys. Chem. B \textbf{109}, 21942 (2005).

\bibitem{Bianchi06} E. Bianchi, J. Largo, P. Tartaglia, E. Zaccarelli, and
F. Sciortino, \ Phys. Rev. Lett. \textbf{97}, 168301 (2006).

\bibitem{Michele06} C. De Michele, S. Gabrielli, P. Tartaglia, and F.
Sciortino, J. Phys. Chem. B \textbf{110}, 8064 (2006).

\bibitem{Lomakin99} A. Lomakin, N. Asherie, and G. B. Benedek, Proc. Natl.
Acad. Sci. USA \textbf{96}, 9645 (1999).

\bibitem{Starr03} F. W. Starr, and J. F. Douglas, J. Chem. Phys. \textbf{119}%
, 1777 (2003).

\bibitem{Zhang04} Z. Zhang, and S. C. Glotzer, Nano Lett. \textbf{4}, 1407
(2004).

\bibitem{Glotzer04a} S. C. Glotzer, Science \textbf{306}, 419 (2004).

\bibitem{Glotzer04b} S. C. Glotzer, M. J. Solomon, and N. A. Kotov, AIChE
Journal \textbf{50}, 2978 (2004).

\bibitem{Sciortino07} F. Sciortino, E. Bianchi, J. F. Douglas, and P.
Tartaglia, J. Chem. Phys. \textbf{126}, 194903 (2007).

\bibitem{Jackson88} G.\ Jackson, W. G. Chapman, and K. E. Gubbins, Mol.
Phys. \textbf{65}, 1 (1988).

\bibitem{Ghonasgi95} D. Ghonasgi, and W. G. Chapman, J. Chem. Phys. \textbf{%
102}, 2585 (1995).

\bibitem{Sear99} R. P. Sear, J. Chem. Phys. \textbf{111}, 4800 (1999).

\bibitem{Mileva00} E. Mileva, and G. T. Evans, J. Chem. Phys. \textbf{113},
3766 (2000).

\bibitem{Kern03} N. Kern, and D. Frenkel, J. Chem. Phys. \textbf{118}, 9882
(2003).

\bibitem{Zhang05} Z. Zhang, A. S. Keys, T. Chen, and S. C. Glotzer,
Langmuir, \textbf{21}, 11547 (2005).

\bibitem{Wilber06} A. W. Wilber, J. P. K. Doye, A. A. Louis, E. G. Noya, M.
A. Miller, and P. Wong, J. Chem. Phys. \textbf{127}, 085106 (2007).

\bibitem{Doye07} J. P. K. Doye, A. A. Louis, I-C. Lin, L. R. Allen, E. G.
Noya, A. W. Wilber, H. C. Kok, and R. Lyus, Phys. Chem. Chem. Phys. \textbf{9%
}, 2197 (2007).

\bibitem{Cummings86} P. T. Cummings, and L. Blum, J. Chem. Phys. \textbf{84}%
, 1833 (1986).

\bibitem{Wei88} D. Wei, and L. Blum, J. Chem. Phys. \textbf{89}, 1091 (1988).

\bibitem{Blum90} L. Blum, P. T. Cummings, and D. Bratko, J. Chem. Phys. 
\textbf{92}, 3741 (1990).

\bibitem{Blum71} L. Blum and A. J. Torruella, J. Chem. Phys. \textbf{56},
303 (1971).

\bibitem{Protsykevich03} I. A. Protsykevich, Condensed Matter Phys. \textbf{6%
}, 629 (2003).

\bibitem{note1} Recall that the Dirac delta function is defined by $%
\int_{a}^{b}\delta \left( x-x_{0}\right) F(x)dx=F(x_{0})$ \ if \ \ $a\leq
x_{0}\leq b$ \ ( $=0$ \ otherwise), for any\ $F$ continuous at $x=x_{0}$.

\bibitem{Friedman85} H. L. Friedman, \textit{A Course in Statistical
Mechanics }(Prentice-Hall, N. J., 1985).

\bibitem{Lee88} L. L. Lee, \textit{Molecular Thermodynamics of Nonideal
Fluids }(Butterworths, Boston, 1988).

\bibitem{Hansen06} J. P. Hansen, and I. R. McDonald, \textit{Theory of
Simple Liquids, 3rd Ed. }(Academic Press, Amsterdam, 2006).

\bibitem{Gazzillo04} D. Gazzillo, and A. Giacometti, J. Chem. Phys. \textbf{%
120}, 4742 (2004).

\bibitem{Kranendonk88} W. G. T. Kranendonk, and D. Frenkel, Mol. Phys. 
\textbf{64}, 403 (1988).

\bibitem{Miller04} M. A. Miller, and D. Frenkel, J. Phys.: Condens. Matter 
\textbf{16}, S4901 (2004).

\bibitem{Perram75} J. W. Perram, Mol. Phys. \textbf{30}, 1505 (1975).

\bibitem{Frenkel02} D. Frenkel, and B. Smit, \textit{Understanding Molecular
Simulation. From Algorithms to Applications, }p. 221\textit{\ }(Academic
Press, San Diego, 2002).

\bibitem{Gray84} C. G. Gray, and K. E. Gubbins, \textit{Theory of molecular
fluids, Vol. I}, Appendix 3E (Clarendon Press, Oxford, 1984).

\bibitem{Stecki81} J. Stecki, and A. Kloczkowski, Molec. Phys. 
\textbf{42}, 51 (1981).

\bibitem{Chen92} X.S. Chen and F. Forstmann, Molec. Phys. \textbf{76}, 1203 (1992).

\bibitem{Klapp97} S. Klapp and F. Forstmann, J. Chem. Phys. \textbf{106}, 9742 (1997).
\end{thebibliography}
\end{document}